\documentclass[%
 aip,
 amsmath,amssymb,
 reprint,%
]{revtex4-1}

\draft 

\usepackage{graphicx}
\usepackage{dcolumn}
\usepackage{bm}
\usepackage{subcaption}
\usepackage{xcolor}
\usepackage[utf8]{inputenc}
\usepackage[T1]{fontenc}
\usepackage{mathptmx}
\usepackage{etoolbox}
\usepackage{todonotes}
\usepackage{siunitx}

\makeatletter
\def\@email#1#2{%
 \endgroup
 \patchcmd{\titleblock@produce}
  {\frontmatter@RRAPformat}
  {\frontmatter@RRAPformat{\produce@RRAP{*#1\href{mailto:#2}{#2}}}\frontmatter@RRAPformat}
  {}{}
}%
\makeatother

\begin{document}


\title{Optimization of a cosmic muon tomography scanner for cargo border control inspection
} 



\author{Z. Zaher}\email{zahraa.zaher@uclouvain.be}
\affiliation{Centre for Cosmology, Particle Physics and Phenomenology (CP3), Université catholique de Louvain, B-1348 Louvain la Neuve, Belgium}

\author{H. Lay}
\affiliation{Department of Physics and Astronomy, University of Sheffield, Sheffield, S3 7RH, United Kingdom}

\author{T. Dorigo}
\affiliation{ Department of Computer Science, Electrical and Space Engineering, Lule\aa ~University of Technology, 971 87 Lule\aa, Sweden}
\affiliation{INFN, Sezione di Padova, Padova, Italy}
\affiliation{Universal Scientific Education and Research Network, Padova, Italy (https://usern.org/)}

\author{A. Giammanco}
\affiliation{Centre for Cosmology, Particle Physics and Phenomenology (CP3), Université catholique de Louvain, B-1348 Louvain la Neuve, Belgium}

\author{V. Gulik}
\affiliation{Institute of Physics, University of Tartu, W. Ostwaldi 1, 50411 Tartu, Estonia}

\author{C. Hrytsiuk}
\affiliation{Institute of Physics, University of Tartu, W. Ostwaldi 1, 50411 Tartu, Estonia}
\affiliation{GScan OÜ, Mäealuse 2/1, 12618 Tallinn, Estonia}

\author{V. A. Kudryavtsev}
\affiliation{Department of Physics and Astronomy, University of Sheffield, Sheffield, S3 7RH, United Kingdom}

\author{M. Lagrange}
\affiliation{Centre for Cosmology, Particle Physics and Phenomenology (CP3), Université catholique de Louvain, B-1348 Louvain la Neuve, Belgium}

\author{T. Metspalu}
\affiliation{GScan OÜ, Mäealuse 2/1, 12618 Tallinn, Estonia}

\author{G. C. Strong}
\affiliation{MODE Collaboration, Oviedo, Spain (https://mode-collaboration.github.io/)}

\author{C. Turkoglu}
\affiliation{Department of Physics and Astronomy, University of Sheffield, Sheffield, S3 7RH, United Kingdom}

\author{P. Vischia}
\affiliation{Universidad de Oviedo and ICTEA, Oviedo, Spain}


\date{\today}

\begin{abstract}
The past several decades have seen significant advancement in applications using cosmic-ray muons for tomography scanning of unknown objects. One of the most promising developments is the application of this technique in border security for the inspection of cargo inside trucks and sea containers in order to search for hazardous and illicit hidden materials. This work focuses on the optimization studies for a muon tomography system similar to that being developed within the framework of the `SilentBorder' project funded by the EU Horizon 2020 scheme. Current studies are directed toward optimizing the detector module design, following two complementary approaches. The first leverages TomOpt, a Python-based end-to-end software that employs differentiable programming to optimize scattering tomography detector configurations. While TomOpt inherently supports gradient-based optimization, a Bayesian Optimization module is introduced to better handle scenarios with noisy objective functions, particularly in image reconstruction-driven optimization tasks. The second optimization strategy relies on detailed GEANT4-based simulations, which, while more computationally intensive, offer higher physical fidelity. These simulations are also employed to study the impact of incorporating secondary particle information alongside cosmic muons for improved material discrimination. This paper highlights the outcomes and key findings from these optimization studies.
\end{abstract}

\pacs{}

\maketitle 

\section{Introducton}

Muon tomography, also known as muography, is a non-invasive imaging technique that leverages naturally occurring cosmic-ray muons to probe the internal structure of large objects. These muons originate from interactions between high-energy cosmic rays and atomic nuclei in the upper atmosphere, producing showers of secondary particles, primarily pions and kaons. As these mesons decay, they generate muons that travel toward the Earth's surface with minimal energy loss.  At sea level, the flux of cosmic-ray muons is approximately $170~\unit{muons}\cdot\unit{m^{-2}}\cdot\unit{s^{-1}}$, with an average energy of vertical muons of the order $3-4~\unit{GeV}$. Due to their relatively large mass—roughly $200$ times that of an electron - these highly energetic muons have a strong penetrating ability, allowing them to traverse several meters of dense material. This makes them particularly valuable for imaging applications where conventional radiation-based methods, such as X-ray imaging, are ineffective due to limited penetration depth [\onlinecite{Borozdin2003}].\\

Muon tomography can be classified into two main approaches: muon transmission imaging and muon scattering tomography (MST). The transmission method relies on detecting attenuation in muon flux as particles pass through an object, providing insights into its internal composition. This technique is particularly useful for large-scale applications, such as studying volcanic structures [\onlinecite{Alessandro2018}] or uncovering hidden chambers in archaeological sites [\onlinecite{Alvarez1970}]. However, it typically requires prolonged data acquisition periods, which makes it less suitable for applications where rapid imaging is essential. On the other hand, muon scattering tomography (MST) operates by measuring the deflections of muon trajectories caused by multiple Coulomb scattering interactions with the material they pass through. High-atomic-number (Z) materials, such as lead and uranium, tend to cause more significant scattering than lower-Z materials like aluminum or plastic. This property makes MST particularly useful for detecting dense and high-Z substances in diverse applications that span many fields, including nuclear reactor inspection, civil engineering, archaeology, and various industrial and homeland security uses [\onlinecite{IAEA2022}]. In border security applications, MST is a powerful tool for identifying hidden threats and dense contraband in cargo containers, such as radioactive materials or explosives. Furthermore, MST is well-suited for scanning smaller objects and offers the advantage of rapid data acquisition, making it an ideal technology for real-time security applications where both speed and precision are critical for border security operations [\onlinecite{barnes2023cosmic}].\\

As muons traverse a material medium, they primarily interact with material nuclei through multiple Coulomb scatterings, where small-angle deviations are most frequent, while occasional large-angle deflections occur as described by Rutherford scattering. After crossing a macroscopic distance, the overall scattering angle distribution consists of a Gaussian core, which accounts for the majority of deflections due to the central limit theorem, along with a longer tail from less frequent large-angle scatterings. For a muon with momentum $p$ [MeV] passing through a material of radiation length $X_0$ [m] over a distance $x$ [m], the root mean square (RMS) width of the Gaussian component of the scattering angle distribution can be approximated by [\onlinecite{pdg}]:
\begin{equation}  \label{eq:thetaRMS}
    \theta_{RMS} = \frac{13.6~\unit{MeV}}{\beta cp}\sqrt{\frac{x}{X_0}} \, ,
\end{equation}

where $\beta c$ is the muon's velocity. Hence, by analyzing the scattering angles of muons before and after traversing a volume of interest, valuable insights into the material densities within can be obtained, given the inverse relationship between $X_0$ and the material density.\\

Optimizing the design parameters of an MST scanner is crucial to achieving optimal performance, as different configurations impact key factors such as acceptance, detection efficiency, and angular resolution. Design parameters such as the spatial arrangement and size of detectors play a critical role in determining the system’s ability to accurately track muon trajectories, maintain a high signal-to-noise ratio, and effectively classify materials. A suboptimal design may lead to reduced tracking precision, lower detection rates, and decreased overall scanning effectiveness. Prior optimization of MST systems has primarily relied on simulations combined with grid-search approaches to evaluate discrete configurations. However, a systematic, end-to-end optimization strategy that directly links detector design to material discrimination or scanning performance has, to our knowledge, not yet been explored. This motivates the approach presented in this work, which seeks to optimize the scanner design in a holistic, performance-driven manner.\\

In this work, we focus on optimizing the SilentBorder scanner (https://silentborder.eu/), a modular MST system designed for cargo screening applications. The scanner consists of multiple hodoscopes, each of which is a self-contained detector module comprising three equally-spaced sensitive detector panels inside a protective casing. The panels are made of a mat of plastic scintillator fibers 1 mm in diameter with a pitch of 7.5 mm, achieving a spatial resolution of 120 $\unit{\mu m}$ and an angular resolution of a few tenths of a milliradian at meter-scale separations [\onlinecite{barnes2023cosmic}]. Compared to drift tubes scanners- for example, those used at LANL [\onlinecite{Priedhorsky2003}]- which achieve sub-millimeter resolutions but require complex infrastructure including gas handling, high-voltage systems, and bulky mechanical support, scintillating fibers offer a more compact solution at the cost of a high channel density and demanding optical/electronic readout. In this sense, fibers represent a balance between operational complexity and intrinsic spatial granularity. These hodoscopes are arranged around the scanned cargo, forming a tracking system capable of measuring muon trajectories before and after interaction with the inspected volume. Due to its modular design, the SilentBorder scanner offers flexibility in configuration, allowing different setups to be explored for improved detection performance. To systematically optimize the design parameterization of the hodoscopes, we adopt a dual-method approach combining Monte Carlo simulations and differentiable optimization techniques [\onlinecite{MODE2022},\onlinecite{MODE2023}]:

\begin{itemize}
\item \textbf{GEANT4-based simulation framework:} We employ a GEANT4 [\onlinecite{GEANT4}]  simulation framework to model different SilentBorder scanner configurations and various cargo materials. GEANT4, a widely used toolkit for simulating particle interactions with matter, provides a realistic modeling environment for evaluating the performance of different scanner layouts.

\item \textbf{TomOpt-based optimization:} TomOpt [\onlinecite{TomOpt-Strong2023}] is a differentiable simulation framework developed for optimizing MST scanner designs through gradient descent.

\end{itemize}

In Section~\ref{sec:g4}, we present the GEANT4 simulation framework. Using this framework, Section~\ref{sec:gaps} provides an analysis of the impact of hodoscope positioning on the reconstruction, and Section~\ref{sec:secondaries} presents a study on the impact of secondary particle hits on event reconstruction. Section~\ref{sec:tomopt} introduces the TomOpt framework, detailing the reconstruction algorithms used. In Section~\ref{sec:bo}, we extend TomOpt with Bayesian optimization and compare the optimization results between the reconstruction algorithms introduced in Section~\ref{sec:tomopt}. Finally, in Section~\ref{sec:conclusion}, we summarize the main findings of this work and outline future directions for improving the SilentBorder scanner design and reconstruction performance.

\section{GEANT4 Simulation Framework}
\label{sec:g4}
A GEANT4-based framework was developed by the SilentBorder consortium to provide a simplified model of the SilentBorder scanner concept with the flexibility to simulate a variety of different scanner configurations and cargo contents.\\

\begin{figure}[h!]

    \centering
    \begin{subfigure}{\linewidth}
        \centering
        \includegraphics[width=\linewidth]{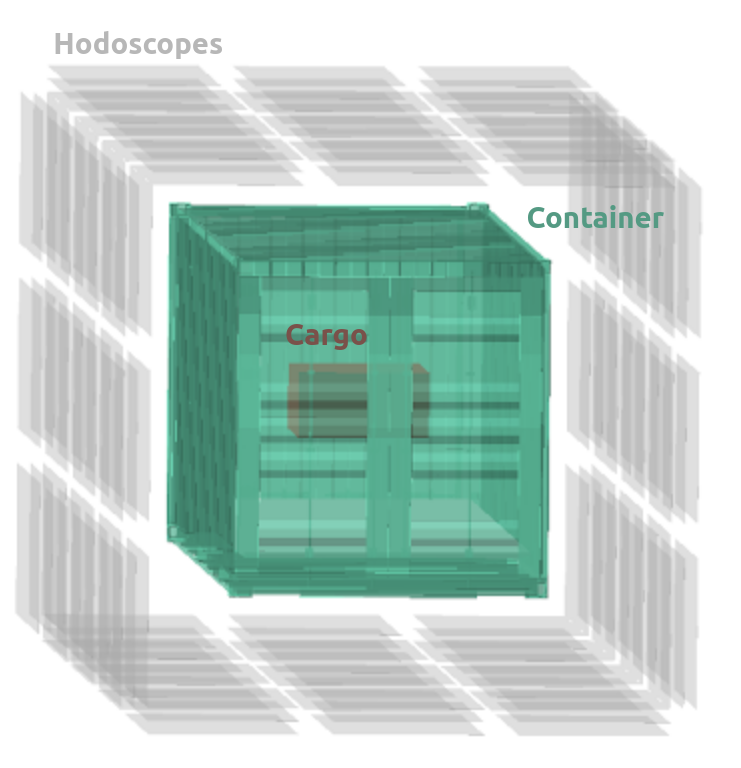}
        \caption{}
        \label{fig:panel_detector}
    \end{subfigure}%
    \\ 
    \begin{subfigure}{\linewidth}
        \centering
        \includegraphics[width=\linewidth]{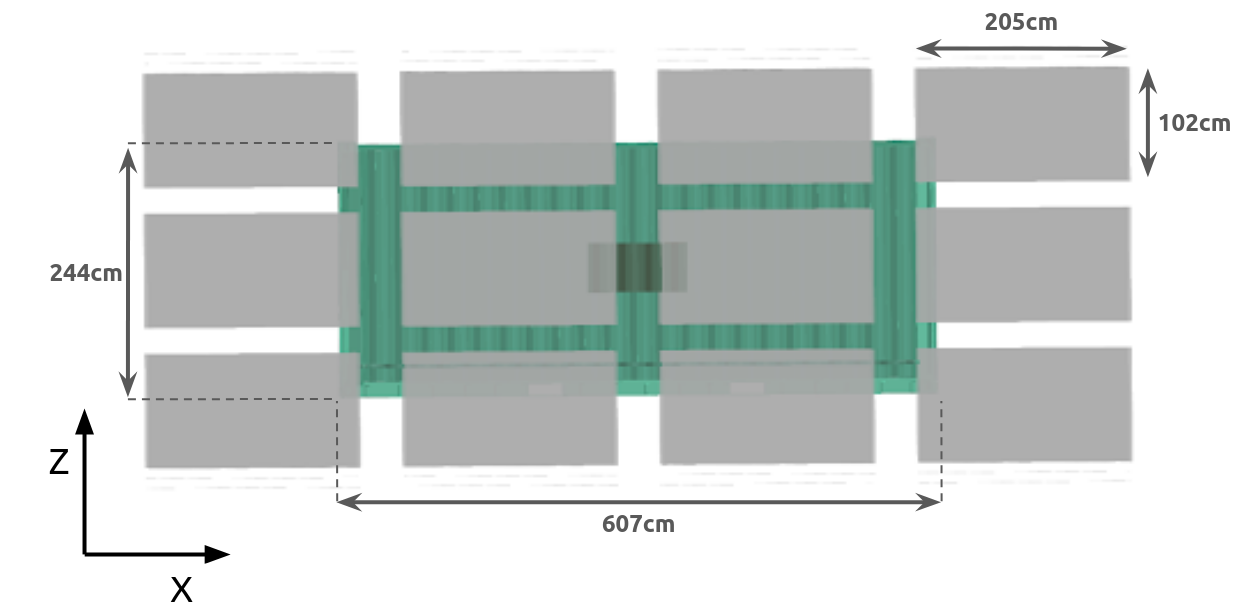}
        \caption{}
        \label{fig:hod_detector}
    \end{subfigure}
    \caption{A visualization of a particular configuration of the detector setup in the GEANT4 simulation. The sensitive detector panels are shown in gray, the shipping container in green and a small cargo object in red. The front view (a) shows that the coverage extends around all four long sides of the container whilst the gaps between hodoscope panels are emphasized in the side view (b).}
    \label{fig:g4detectorvisual}
\end{figure}

Geometry Description Markup Language (GDML) [\onlinecite{gdml}] is used to describe the size, shape, and location of the relevant objects and materials. Firstly, an empty shipping container of external dimensions 6.07\,m$\times$2.44\,m$\times$2.60\,m is placed at the center of the geometry with a plywood floor of depth 40\,cm. The container is surrounded by a total of 48 hodoscopes, a 3$\times$4 grid on each of the long sides of the container. This arrangement is visualized in Fig.~\ref{fig:g4detectorvisual}. Each hodoscope is itself comprised of 3 layers (plates) of solid plastic scintillator each of dimensions 2.1\,m$\times$1.1\,m$\times$3.3\,mm and separated by 10\,cm. Note the casing and electronics for each hodoscope are not simulated in this simplification, but the 10\,cm gaps are introduced to account for the `dead regions' they introduce. The plate is simulated as a solid scintillator rather than as a mat of scintillator fibers. This significantly reduces the complexity of the geometry description and can be considered a good approximation of the fiber mats once the smearing for detector resolution effects is applied later.
\\

A number of different cargo scenarios can be simulated. The setup shown in Fig.~\ref{fig:g4detectorvisual} involves the inclusion of a 1\,m$\times$1\,m$\times$0.5\,m volume in the center of the container. This volume can be simulated using the bulk properties of a number of materials including:

\begin{itemize}
    \item metals, such as iron, steel or lead
    \item explosives, such as TNT, RDX or PETN
    \item illegal drugs, such as cocaine or cigarettes
    \item typical legal cargos, such as bananas or toilet roll
    \item or nuclear materials, such as uranium.
\end{itemize}

Options are also available to fill the entire container with a single cargo, split the container between a number of cargos, or include a single small element of illicit material and fill the remaining space with a legal material.\\

Air shower cosmic-ray muons can be simulated in this volume using the Muons for Silent Border (MuSiBo) generator. MuSiBo utilizes a Gaisser parameterization of the muon flux, with modifications to account for muon decay and the Earth’s curvature. It is based on work presented in reference [\onlinecite{Blackwell_2015}]. Muons are sampled on the surface of a cube, a method which better accounts for the high-angle flux when compared to sampling on a single surface above the region of interest. For this simulation, a box of 11\,m$\times$5\,m$\times$4\,m was used to fully enclose the volume of the container and all of the hodoscopes.\\

Finally, the initial muons simulated with MuSiBo are propagated through the simulated geometry via use of the GEANT4 software package [\onlinecite{GEANT4}]. Specifically, the \texttt{QGSP\_BERT} physics list is utilized in version 11.0.p03 of the software. Muons, and any secondary particles produced, are stepped through the full geometry with information saved about their trajectory and total energy deposition within each logical volume.\\

\section{GEANT4 Study of Hodoscope Positioning}
\label{sec:gaps}
The relative positioning of hodoscope plates can impact the performance of the scanner system. This impact can be appreciated in both the efficiency in collecting ‘interesting’ muons and the angular resolution of their trajectories.\\

A study is performed using the GEANT4 simulation framework described in Section~\ref{sec:g4}. For the rest of this section we will use a coordinate system that is relative to each hodoscope group; $x$ will refer to the long in-plane dimension, $y$ to the short in-plane dimension, and $z$ to the out-of-plane dimension. This ensures that moving hodoscope plates in one coordinate is conceptually the same regardless of the rotational orientation of the hodoscope group (top, bottom or sides).
\\

A baseline configuration was established with the container filled entirely with wood and the hodoscopes as close together in $x$ and $y$ as possible. Note that due to space reserved for electronics and casing this still leaves a $\sim$10\,cm gap between the active scintillator areas of adjacent hodoscopes. The default spacing in $z$ between the three plates that form a single hodoscope is 10\,cm.\\

A number of scenarios are created by independently varying the spacing in $x$, $y$ and $z$. Due to physical constraints (such as not overlapping any hodoscope regions) the variations are different for each direction. The cargo is kept identical and for each scenario a 15 minute exposure to the cosmic flux is provided by the MuSiBo generator. This 15 minute exposure corresponds to a total of 12,683,700 muons in the volume used by MuSiBo, and the GEANT4 simulation required an average of 14.5 CPU hours to propagate and save information for each of these muons.\\

The energy depositions recorded by GEANT4 in the active hodoscope plates pass through a pseudo detector simulation. At this stage, in order to reduce computational time, we only consider events with 2 or more depositions before and 2 or more depositions after the cargo. Without this requirement a muon will not be reconstructed later on, so we can save resources by discarding this event here. This step is later referred to as `filtering'. Any depositions recorded within 1\,mm and 20\,ns of each other are combined and a Gaussian positional smearing, with width 1\,mm, is then applied to their computed barycenter. This smearing, larger than the real intrinsic per-plate resolution of 120\,$\mu$m, is applied to conservatively approximate the combined effects of clustering, optical cross-talk, and readout not explicitly modeled in GEANT4. Finally, a threshold of 0.4\,MeV is applied to ensure the deposition would have created a detector ‘hit’.\\

These hits form the input to a simple reconstruction chain. Tracks are formed when one hit is found on at least two hodoscope plates. These two plates must be found in different layers of the same hodoscope group (above, below, left or right). If a hodoscope plate has received multiple hits it is excluded from track building. This only impacts about 3.5\% of track candidates, and therefore the small reduction in efficiency is worth accepting for the improved angular resolution. Whilst a more sophisticated algorithm could be designed to make use of this subset, the reduction in operating times for field-based MST scanners would be minimal.
\\

\begin{figure}[h!]
    \centering
    \includegraphics[width=.9\linewidth]{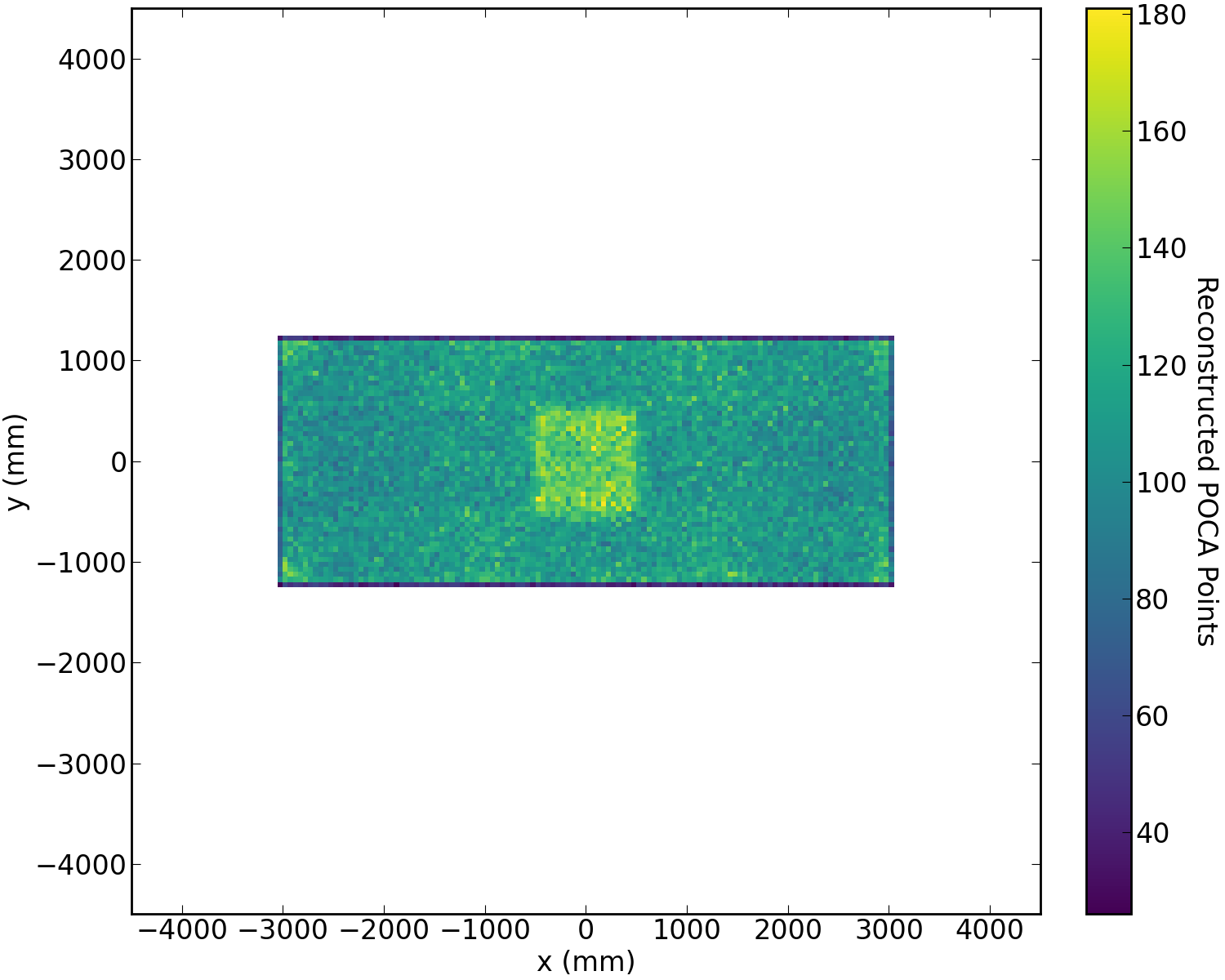}
    \caption{Locations of reconstructed PoCA points within the container in the $xy$-plane (looking down on the container). The scenario used a 100$\times$100$\times$50\,cm cargo of lead, centered in the container.}
    \label{fig:poca_example}
\end{figure}

Linear regression is used to find the best fit line between the hits providing the direction of the reconstructed muon track. The final stage of the reconstruction takes events in which the muon had tracks reconstructed in two different hodoscope groups. A simple Point-of-Closest-Approach (PoCA) reconstruction is performed from which the location and angle of the scatter point is determined. An example of the output of the PoCA reconstruction is shown in Fig.~\ref{fig:poca_example}. The reconstruction chain is significantly less computationally expensive than the simulation, each scenario requiring an average of 30 CPU minutes to perform the full reconstruction and save the output.\\

A number of metrics are used to quantify the change in performance when the hodoscope positions are adjusted. They can be split into two categories.\\

The first three metrics assess the efficiency of the system. As a proportion of the total number of muons simulated by MuSiBo (12,683,700) we can evaluate:
\begin{itemize}
    \item The number of events with a `Muon of Interest' defined as the number of muons that make any deposition in any of the hodoscopes' scintillator or the cargo of interest.
    \item The number of events in which a PoCA point was reconstructed.
    \item The number of events in which a PoCA point was reconstructed inside the container - thus providing information about the cargo.
\end{itemize}

\begin{figure}[h!]
    \centering
    \begin{subfigure}{\linewidth}
        \includegraphics[width=.9\linewidth]{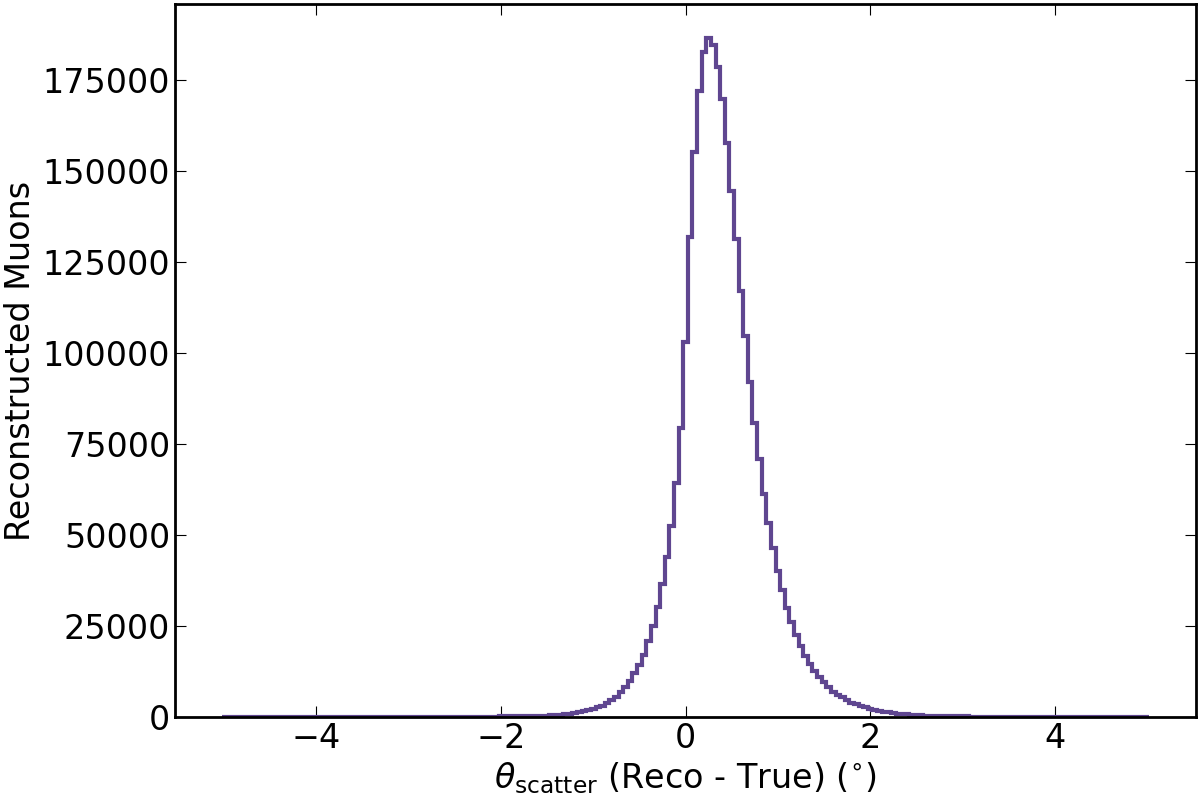}
        \label{fig:scat_angle_nominal}
        \caption{}
    \end{subfigure}
    \begin{subfigure}{\linewidth}
        \includegraphics[width=.9\linewidth]{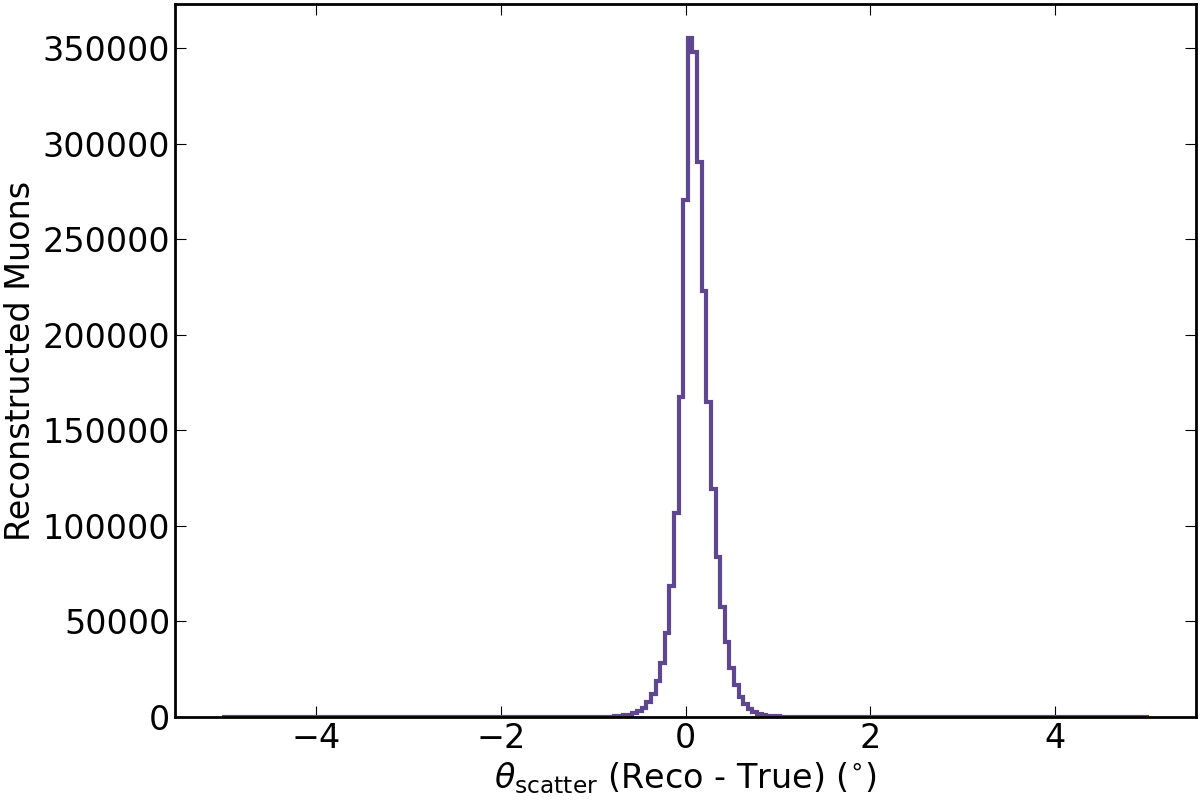}
        \label{fig:scat_angle_increased_spacing}
        \caption{}
    \end{subfigure}
    \caption{The resolution of the reconstructed scattering angle in two different example scenarios, (a) the nominal configuration and (b) where the vertical spacing between hodoscope plates has been increased.}
    \label{fig:ang_res_example}
\end{figure}

The other two metrics assess the angular reconstruction of the muon scattering point. The difference between the reconstructed and true scattering angles is asymmetric due to the positive-only definition of a scattering angle. For smaller angles, the reconstructed angle is more likely to be larger in magnitude than the true angle. Fig.~\ref{fig:ang_res_example} shows the resolution in two different example scenarios. We can calculate the mean and standard deviation of this difference distributions as metrics to use in the variation studies.\\

\begin{figure}[h!]
    \centering
    \includegraphics[width=.9\linewidth]{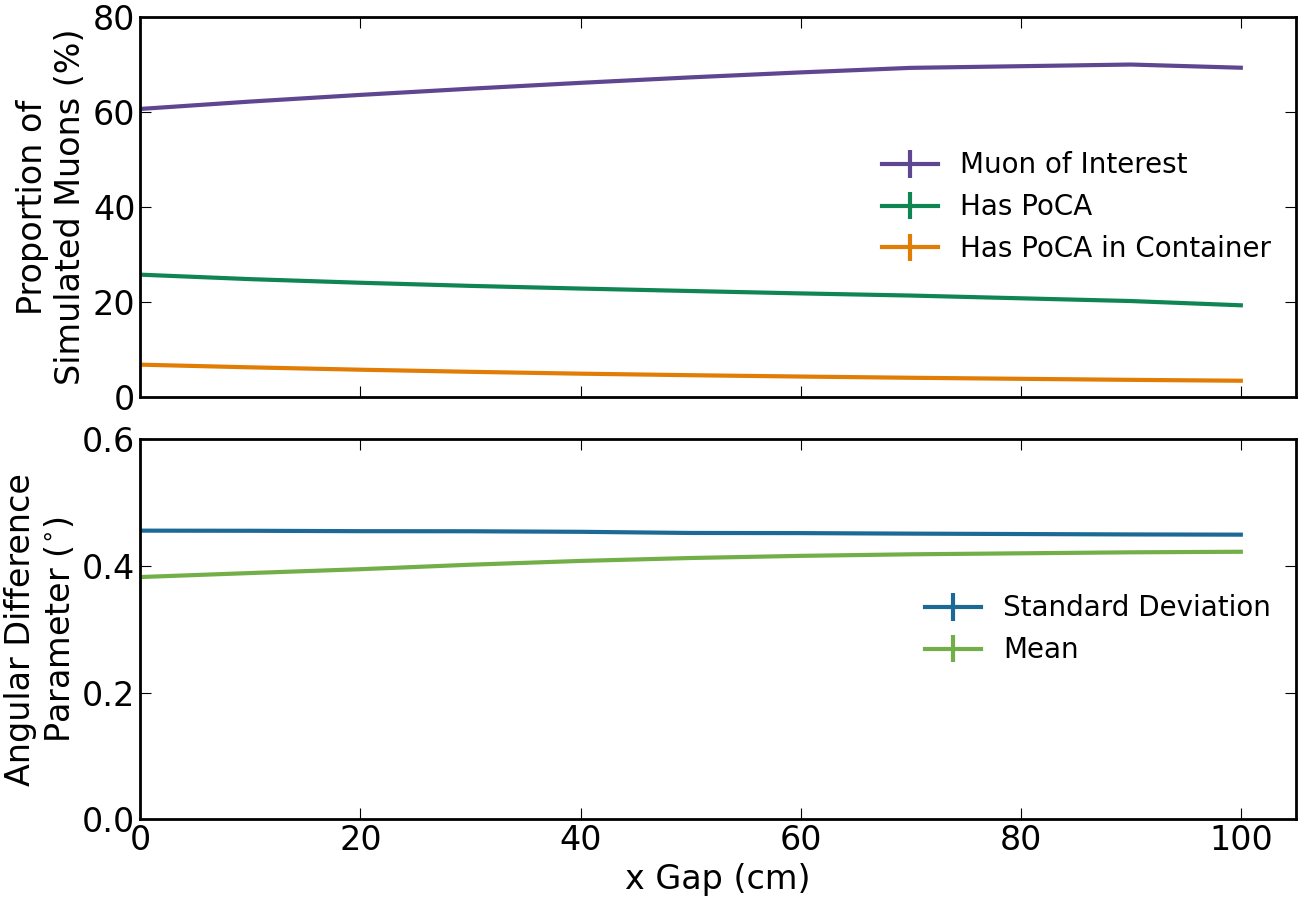}
    \caption{The change in efficiency (top) and angular resolution (bottom) metrics as the gap between hodoscopes in the $x$-direction is varied.}
    \label{fig:x_gaps}
\end{figure}

Fig.~\ref{fig:x_gaps} shows the variation in the efficiency and angular resolution metrics as a function of the gaps between the hodoscopes in $x$. Note that 0\,cm corresponds to the minimum gap between the hodoscope objects. However, a $\sim$10\,cm gap between the active scintillator will still be present due to the casing and electronics space reserved on each hodoscope. The behavior is identical for $y$-variations due to the fact that these variations are both in the plane of the hodoscope plates. Due to physical limitations, however, there is a much larger range of motion in the x-direction (the length of the container) than the y-direction (the width of the container).\\

In both directions the number of muons of interest increase as the gaps increase, this is purely because the hodoscopes as an ensemble now cover a larger volume within the total simulation world and hence more of the muons are likely to intercept at least one hodoscope. However, the number of PoCA points, and specifically those in the cargo, decreases due to the gaps that are created in the regions closest to the cargo itself. There is a subsequent reduction in the number of muons that intercept hodoscopes before and after the container and pass through the cargo. The reduction in useful PoCA points per muon will mean that a longer exposure time is required in order to achieve the same level of sensitivity.\\

There is no visible change in the standard deviation of the angular difference metric, so the angular resolution does not improve. There is a small increase in the mean when going to the largest gaps. This is likely due to the reduced available phase space in the scattering angle created by the gaps.\\

\begin{figure}[h!]
    \centering
    \includegraphics[width=.9\linewidth]{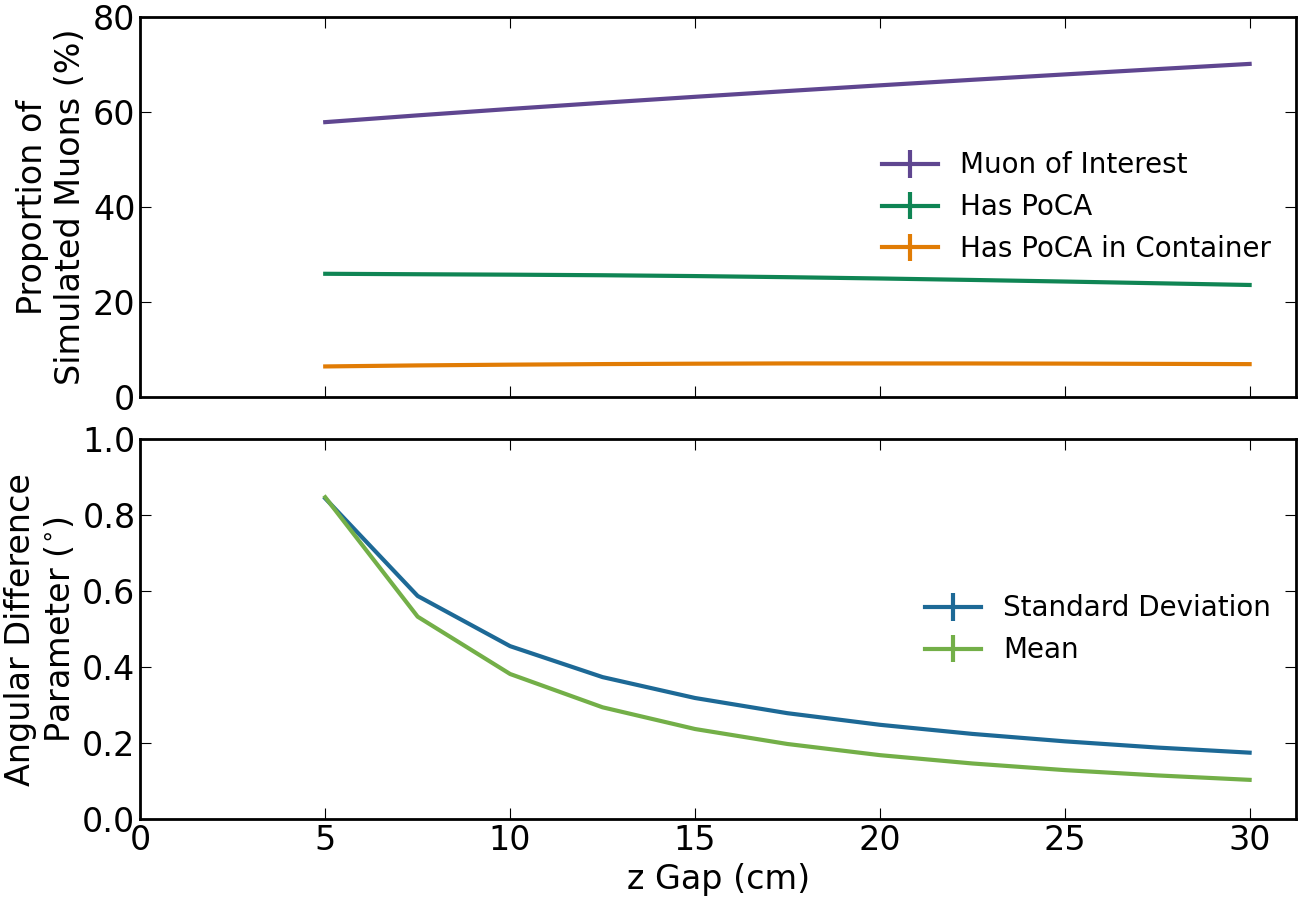}
    \caption{The change in efficiency (top) and angular resolution (bottom) metrics as the gap between hodoscopes in the $z$-direction is varied.}
    \label{fig:z_gaps}
\end{figure}

Fig.~\ref{fig:z_gaps} shows the variation in efficiency and angular resolution metrics as a function of the vertical spacing between the hodoscope plates in $z$. The nominal gap between the plates is 10\,cm. As with the $x$ variations, the number of interesting muons increases as the total scanner volume increases with larger spacing. The number of PoCA points slightly decreases. This occurs because the probability of the muon hitting enough plates to ensure track reconstruction before and after the cargo decreases, but the effect is very minimal. The angular resolution on the other hand is dramatically impacted by the widening of the vertical spacing. By increasing the spacing both the standard deviation and mean parameters reduce significantly before beginning to flatten out. As the distance between the two or three detection points increases the positional resolution of each plate has a smaller impact on the reconstructed angle, hence improving the angular resolution. The reduction in the mean occurs due to the asymmetry of the distribution. As the resolution improves the asymmetry is naturally reduced in magnitude.\\

These variation studies can be used to inform decisions about the design for a full-scale SilentBorder-like scanner system. As could be anticipated, minimizing the in-plane gaps between hodoscopes improves the efficiency of the system, thus reducing the necessary exposure time and making the system quicker to use and more viable in border security. Increasing the vertical out-of-plane spacing between hodoscope plates improves the angular resolution due to the lever arm effect, although it does so at the expense of efficiency. For a particular system an appropriate trade-off balance can be found, in the scenario examined in this study Fig.~\ref{fig:z_gaps} suggests that up to $\sim$20\,cm the increase in out-of-plane spacing improves the angular resolution enough for the reduction in efficiency to be worthwhile. Beyond this, the resolution improvements plateau and therefore would be unlikely to be worth the efficiency reduction. Note that these studies were completed with a simple PoCA reconstruction and decisions were made to ignore plates with multiple hits or events with 3 or more tracks. These considerations will have an impact on the magnitude of the effects presented but should not undermine the trajectories of the changes and thus the conclusions drawn.

\section{GEANT4 Study of Secondary Hits}
\label{sec:secondaries}
As muons pass through the hodoscope, cargo, container, and surrounding air they can produce secondary particles, primarily delta rays. An assessment is made in this section of the impact of any hodoscope hits induced by such secondary activity. Note that activity due to other particles generated in cosmic air showers is not considered in this study but can also be considered as a source of `secondary' information in MST. \\

The GEANT4 simulation described in Section~\ref{sec:g4} shows that about 11\,\% of the energy depositions recorded in the hodoscope plates resulted from non-muons. However, many of these depositions would likely be irresolvable from the accompanying muon hits with the SilentBorder scanner setup. Using approximate numbers provided by the GScan team, a pseudo detector simulation was performed (same as Section~\ref{sec:gaps}) in which depositions within the same 1\,mm$\times$1\,mm area and 20\,ns of each other were combined into a single hit. For these amalgamated hits all the relevant energy depositions were summed. Following this amalgamation procedure 8.3\,\% of the remaining hits resulted purely from non-muon activity, with the remaining 91.7\% having at least some contribution from the muon.\\

\begin{figure}[h!]
    \centering
    \includegraphics[width=\linewidth]{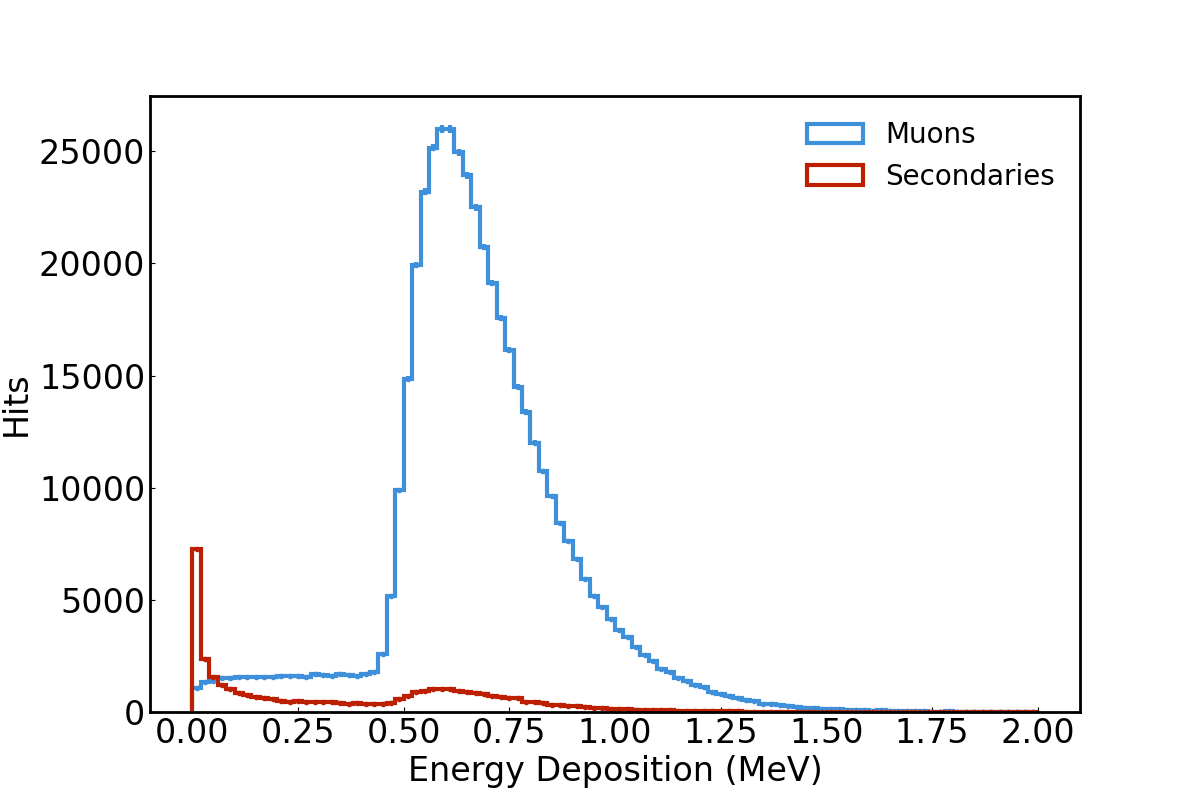}
    \caption{The total true energy deposition for each hodoscope plate hit following the application of detector resolution constraints. The distribution is split into hits with some contribution from the primary muon (blue) and hits with no contribution from the primary muon (red).}
    \label{fig:secondariesedep}
\end{figure}

Fig.~\ref{fig:secondariesedep} shows the energy deposited per hit (in a single hodoscope plate) for these two categories of hits. The energy deposited by a minimally ionizing particle  (MIP) crossing the full ~3.3\,mm plate depth is around 0.6\,MeV, and a peak is seen in this region for both muon and secondary hits. The flat shelf at lower energy depositions is created by particles ``clipping'' a hodoscope plate rather than passing fully through it. The secondaries distribution has a third feature, another peak at very low depositions. This results from the full containment of secondaries produced with very low kinetic energy which only travel a very short distance in the scintillator before coming to a stop.\\

The goal of this study was to evaluate what impact deposits from secondaries could have in degrading the performance of MST with the SilentBorder scanners. It is clear from Fig.~\ref{fig:secondariesedep} that above a certain energy threshold there is no method to distinguish between secondary and muon hits based on the scale of the energy deposition.\\

\begin{figure}[h!]
    \centering
    \begin{subfigure}{\linewidth}
        \centering
        \includegraphics[width=\linewidth]{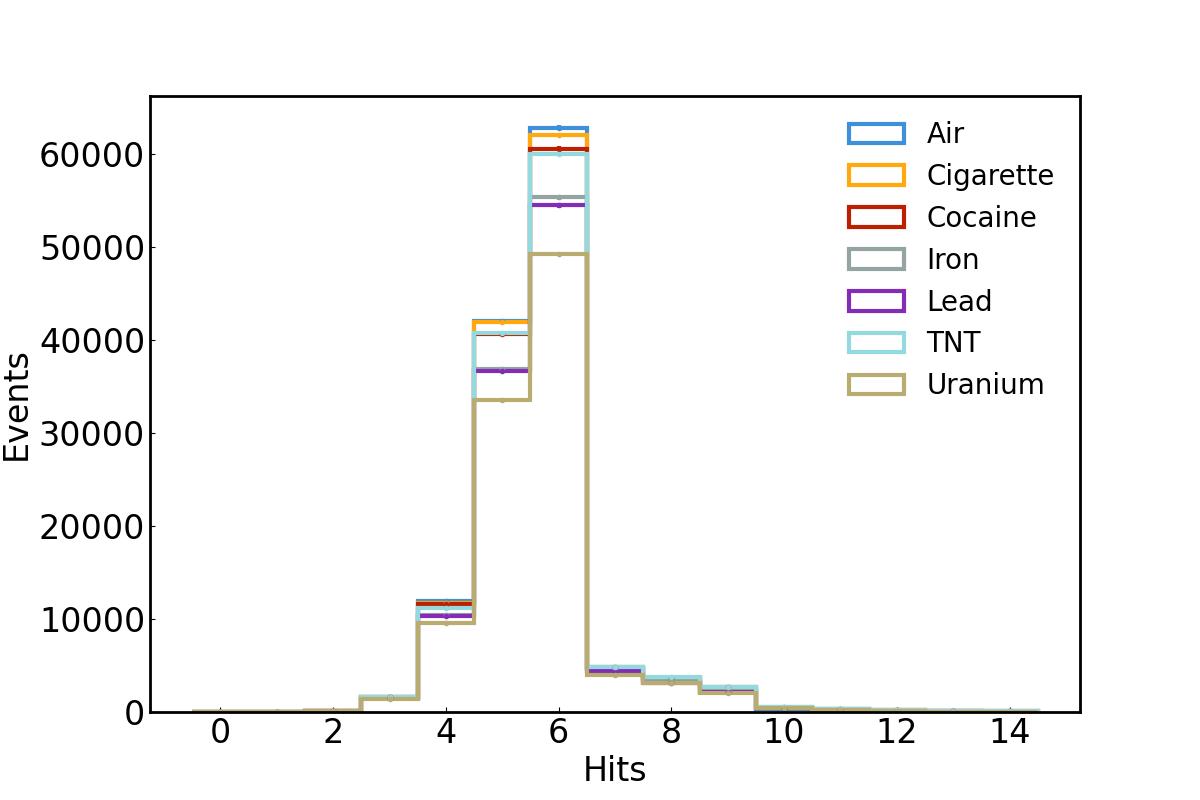}
        \caption{}
        \label{fig:totalhodohits}
    \end{subfigure}
    \begin{subfigure}{\linewidth}
        \centering
        \includegraphics[width=\linewidth]{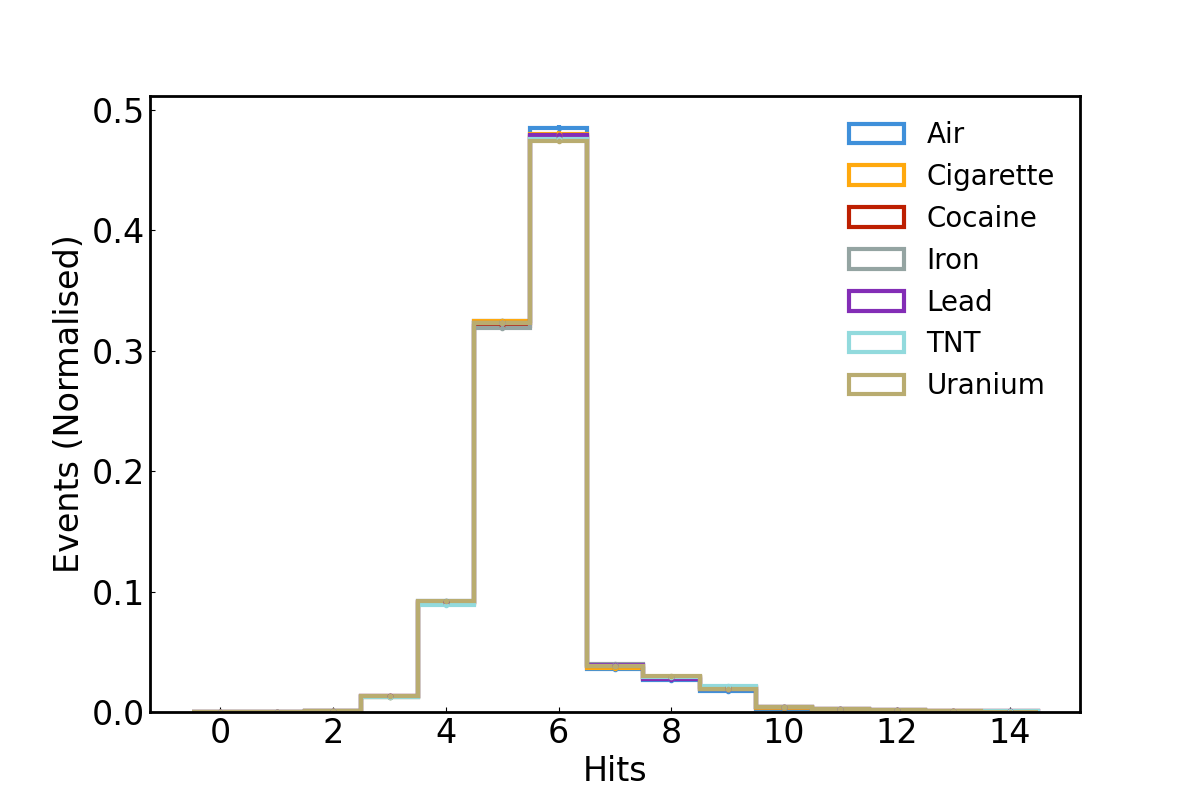}
        \caption{}
        \label{fig:totalhodohitsnorm}
    \end{subfigure}
    \caption{The total number of hits recorded across all hodoscopes for a variety of cargo types (each simulated as a 1\,m$\times$1\,m$\times$0.5\,m box centered in the container) with an otherwise identical configuration. (a) shows this distribution without normalization, whilst (b) shows this distribution after normalizing for the number of events that pass the filtering stage.}
    \label{fig:hodohits}
\end{figure}

For the first study a threshold of 0.4\,MeV is applied, retaining only those depositions that lie within the MIP peak region. A cargo block of size 1\,m$\times$1\,m$\times$0.5\,m is placed at the center of the container with a variety of material types considered. The MuSiBo generator is used to provide muons corresponding to a 15 minute exposure time. Fig.~\ref{fig:totalhodohits} shows the total number of hits recorded across all hodoscopes per event, regardless of whether they were produced via secondaries or muons. In general the peak structure around 6 hits is produced via the 3 hits left by the muon, before and after the cargo whilst the small tail is produced by the presence of any secondary hits. Whilst it appears that there is some variation between the different cargo types this variation stems from the number of events passing the filter, not from a significant difference in the number of secondary hits produced per muon. Higher density and higher-Z materials reduce the number of events that pass the filtering process. Fig.~\ref{fig:totalhodohitsnorm} shows the same distribution but normalized by the number of events that passed the filtering, very limited variation per cargo is now observed. \\

\begin{figure}[h!]
    \centering
    \includegraphics[width=\linewidth]{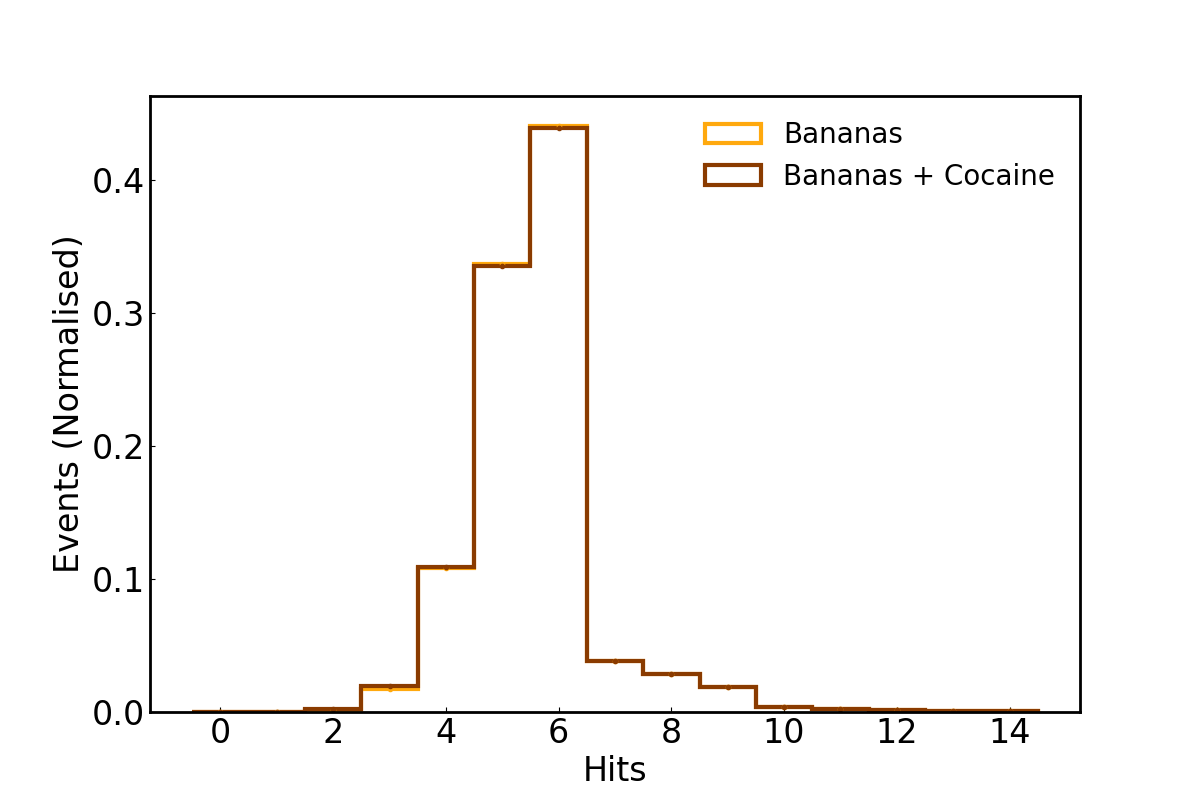}
    \caption{The total number of hits recorded across all hodoscopes normalized for the number of events that pass a filtering stage. Two cargo scenarios are shown, in the first the container is filled with bananas, in the second a 40\,cm$\times$40\,cm$\times$40\,cm box of cocaine is placed in the center of the bananas.}
    \label{fig:totalhodohitsbanana}
\end{figure}

Fig.~\ref{fig:totalhodohitsbanana} shows the same analysis performed for a more `realistic’ smuggling scenario. For this scenario the container is filled with bananas and then a 40\,cm$\times$40\,cm$\times$40\,cm box of cocaine is placed in the center of the bananas. This is compared to the scenario in which only the bananas are stored. Again no significant variation is seen in the number of hits recorded indicating the presence of secondary hits cannot, in and of itself, be used to discriminate between different cargoes. \\

\begin{figure}[h!]
    \centering
    \begin{subfigure}{.8\linewidth}
        \centering
        \includegraphics[width=\linewidth]{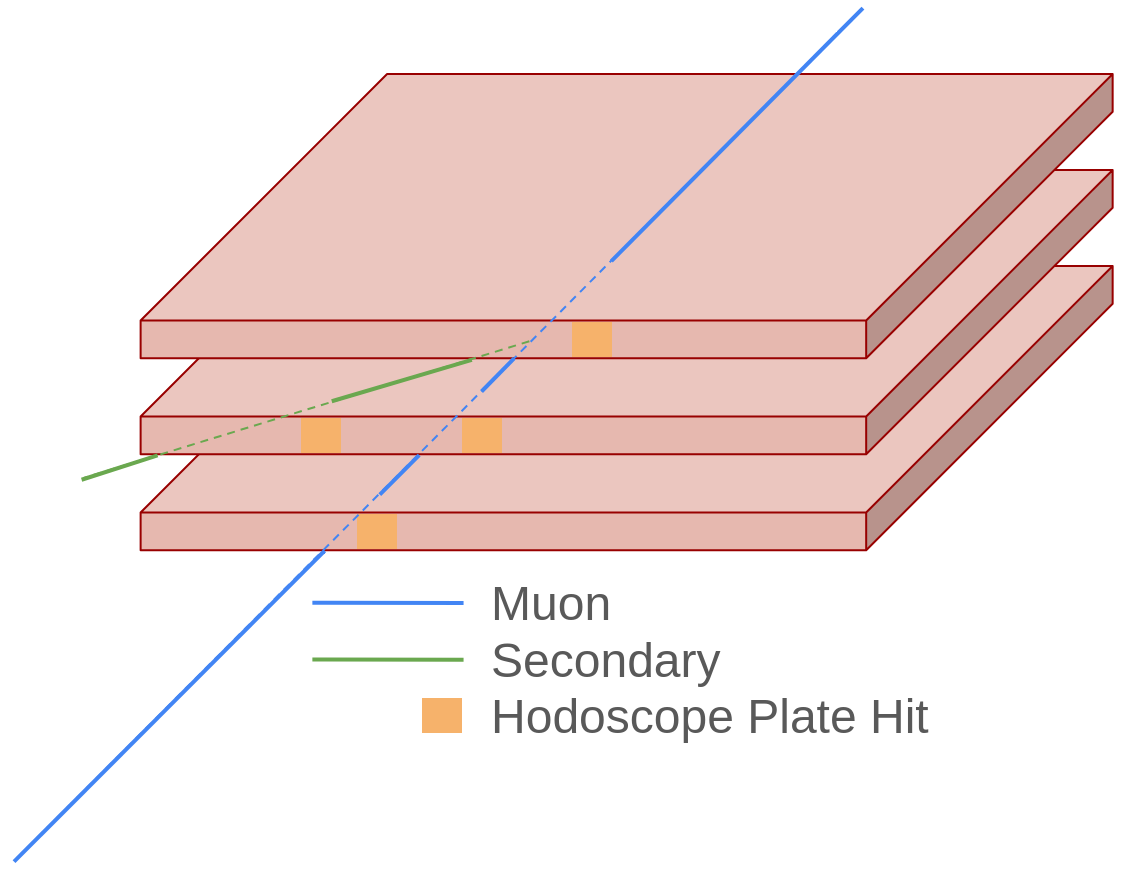}
        \caption{}
        \label{fig:confdiag1}
    \end{subfigure}
    \begin{subfigure}{.8\linewidth}
        \centering
        \includegraphics[width=\linewidth]{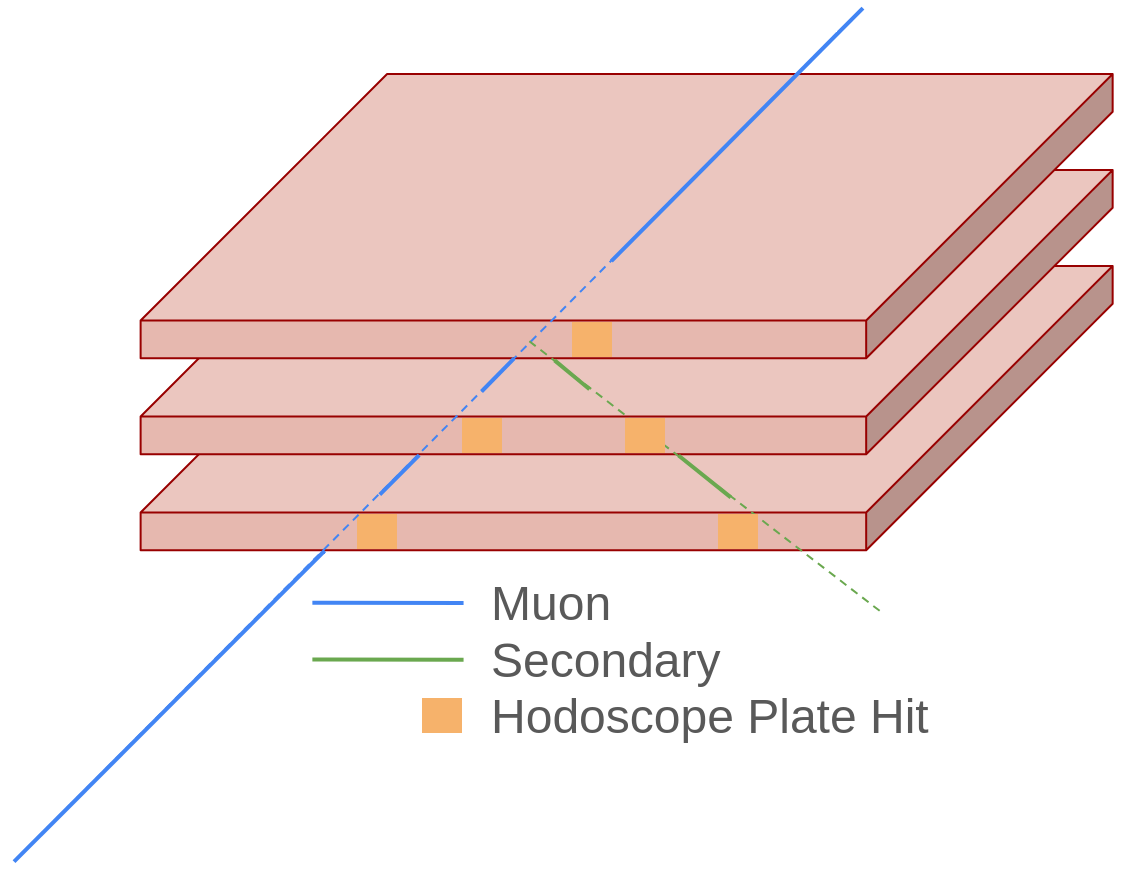}
        \caption{}
        \label{fig:confdiag2}
    \end{subfigure}
    \caption{Two schematic examples of a muon passing through the three plates in a single hodoscope. In (a) the emitted secondary produces just a single hit and the muon trajectory is still clear, in (b) the emitted secondary produces two hits and the muon track is much harder to discern.}
    \label{fig:secondaryconfdiags}
\end{figure}

The next question of interest is whether these hits degrade the usefulness of the muon hits. This can be evaluated by assessing the number of hits within a single hodoscope that result from muons and secondaries respectively within each event. The default assumption is that three hits from a muon, one per hodoscope plate, can be used to produce a muon track direction through that hodoscope. As is illustrated in Fig.~\ref{fig:secondaryconfdiags}, the presence of more than one secondary hit can limit the success of this approach, as identifying which hits correspond to the muon track becomes more complicated. Note, these schematics make two assumptions that are not always true. Firstly, that just one secondary results in the production of hits in a single hodoscope and secondly that the secondary is produced by the muon whilst passing through the hodoscope (as oppose to the container or cargo). These are simplifications for the purpose of visualization, all such situations are considered in the analysis.\\

\begin{figure}[h!]
    \centering
    \includegraphics[width=\linewidth]{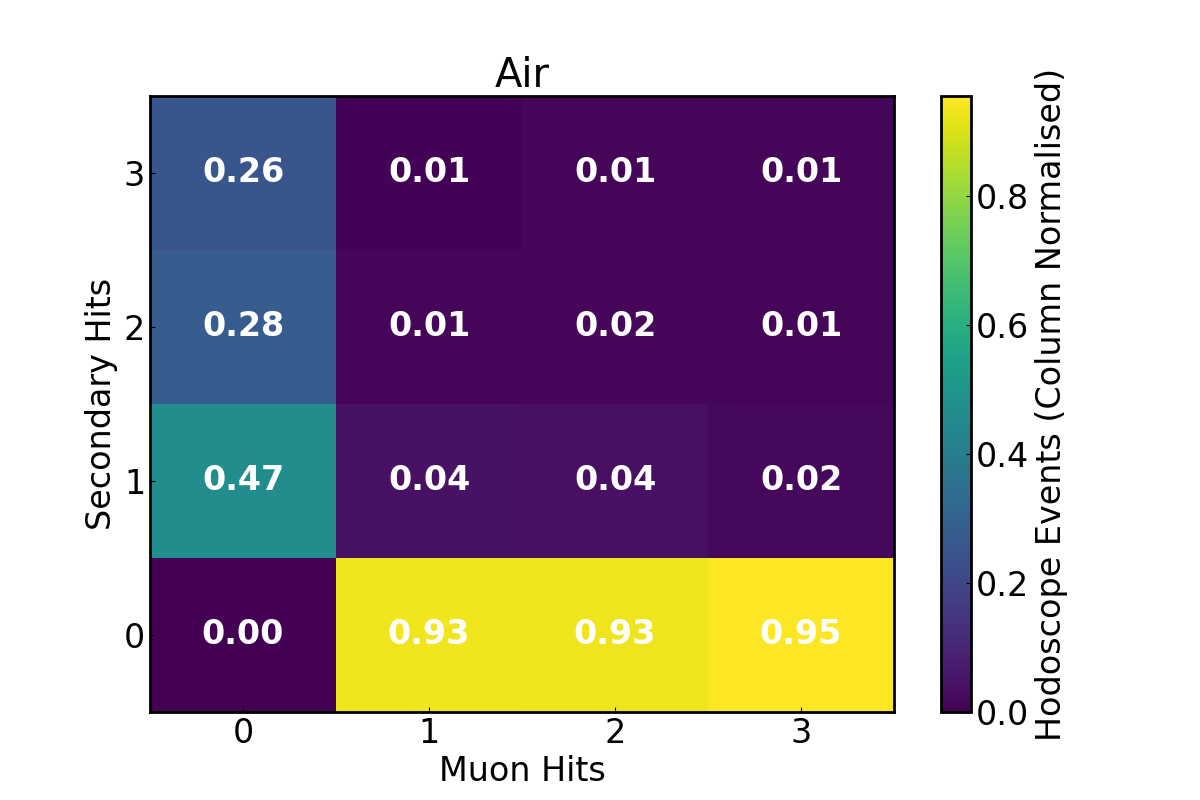}
    \caption{The number of hits (split into muon hits and secondary hits) recorded within a single hodoscope for a geometry containing an empty container. The matrix is column normalized, such that the values represent the proportion of times $y$ secondary hits are recorded given that $x$ muon hits are recorded.}
    \label{fig:confmat_air}
\end{figure}

A confusion matrix visualizes this impact in Fig.~\ref{fig:confmat_air}. The matrix is column normalized such that the values represent the proportion of times in which $y$ secondary hits are recorded given that $x$ muon hits are recorded. In this version of the figure the container is empty of cargo (the cargo is `Air'). We see that in 95\,\% of occasions in which 3 muon hits are recorded in a single hodoscope, no secondary hits are recorded. A further 2\,\% of those occasions have just a single secondary hit, which still allows for identification of this hit as the outlier. Therefore we categorize 3\,\% of the occasions with 3 muon hits as having potential for muon tracking degradation due to the presence of secondary hits. If this value is kept low then such contaminated events can just be removed from consideration without significant impact to the scanner’s sensitivity.\\

We express the contamination proportion (CP) as

\begin{equation}
    \small
    \textrm{CP} = \frac{\textrm{Hodoscopes with 3 muon hits and $\geq$2 secondary hits}}{\textrm{Hodoscopes with 3 muon hits}}.
\end{equation}

\begin{figure}[h!]
    \centering
    \includegraphics[width=\linewidth]{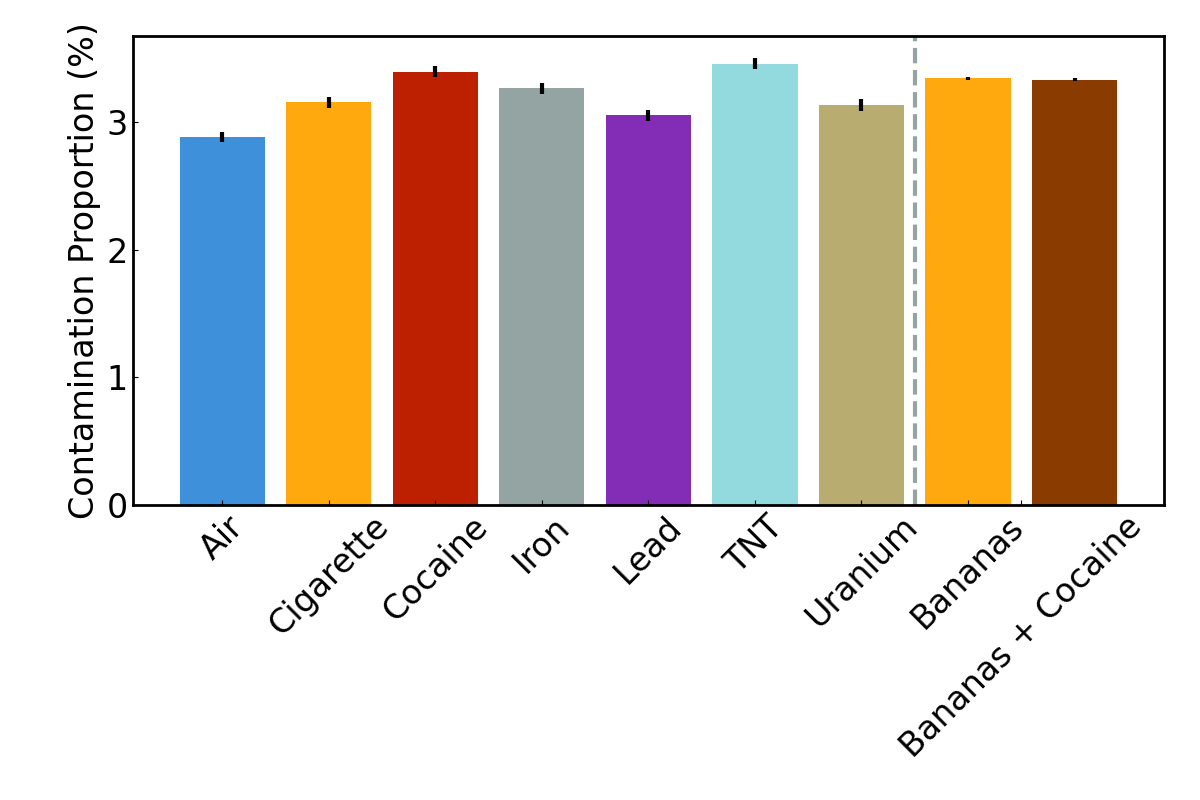}
    \caption{The contamination proportion for a variety of cargo types. The entries on the left of the line show different cargoes each simulated as a 1\,m$\times$1\,m$\times$0.5\,m box centered in the container. The entries on the right of the dashed line show the more `realistic' scenarios where first a container is filled with bananas and then in the second scenario a 40\,cm$\times$40\,cm$\times$40\,cm box of cocaine is placed in the center of the bananas.}
    \label{fig:conf_cargos}
\end{figure}

\begin{figure}[h!]
    \centering
    \includegraphics[width=\linewidth]{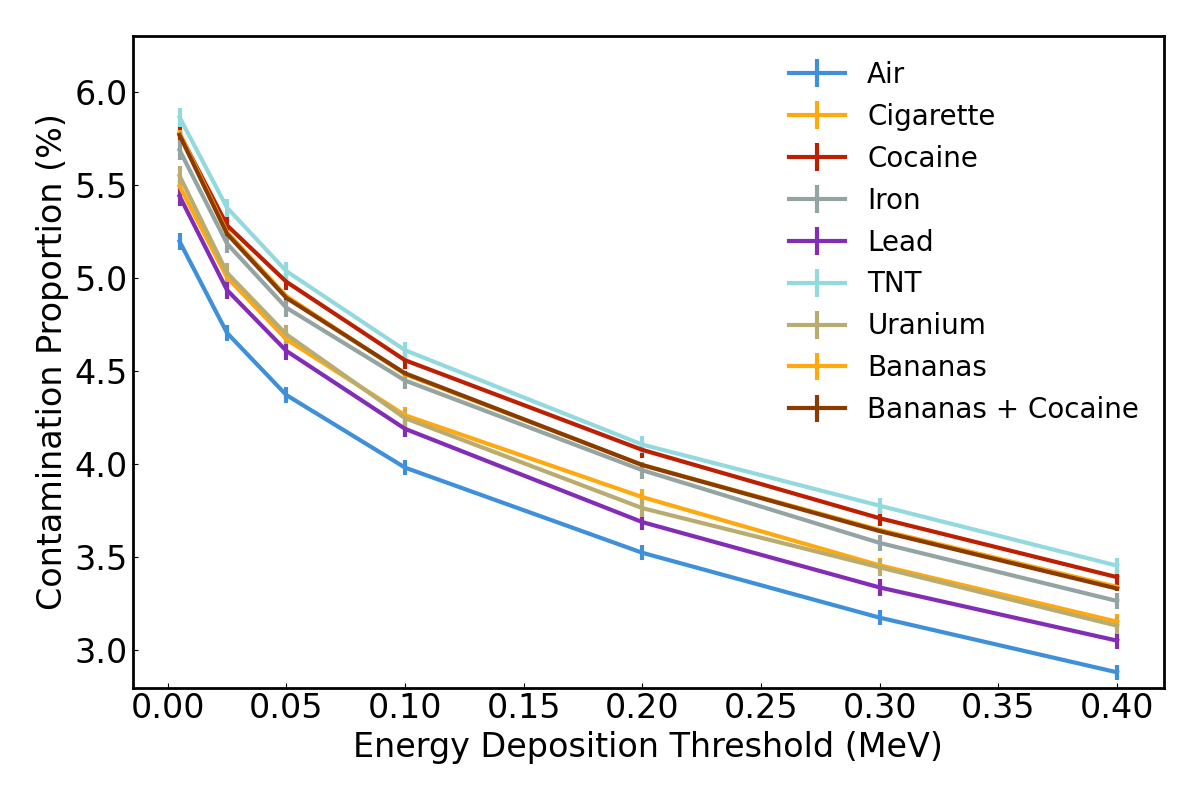}
    \caption{The dependence of the contamination proportion on the energy threshold for a variety of cargo types (each simulated as a 1\,m$\times$1\,m$\times$0.5\,m box centered in the container) with an otherwise identical configuration.}
    \label{fig:conf_thresh}
\end{figure}

We can study how this number varies in a number of different scenarios. Fig.~\ref{fig:conf_cargos} shows that the contamination proportion is unaffected when we vary among the different cargo varieties. As with previous studies to this point, a threshold of 0.4\,MeV was used, with all lower energy hits assumed to be unobservable. Fig.~\ref{fig:conf_thresh} shows how the contamination proportion varies with respect to this threshold. It is clear that as the threshold is reduced the contamination effect increases, as would be expected given the distributions presented in Fig.~\ref{fig:secondariesedep}. However, the effect remains $<6\,\%$ for all scenarios and is therefore unlikely to drastically affect the SilentBorder scanner performance even with a much lower sensitive threshold.\\

Finally, we consider the scenario in which the hodoscope is produced with just two plates not three. In this scenario, just a single secondary hit will result in a confusion effect and so the metric becomes

\begin{equation}
    \small
    \textrm{CP} = \frac{\textrm{Hodoscopes with 2 muon hits and $\geq$1 secondary hits}}{\textrm{Hodoscopes with 2 muon hits}}.
\end{equation}

\begin{figure}[h!]
    \centering
    \includegraphics[width=\linewidth]{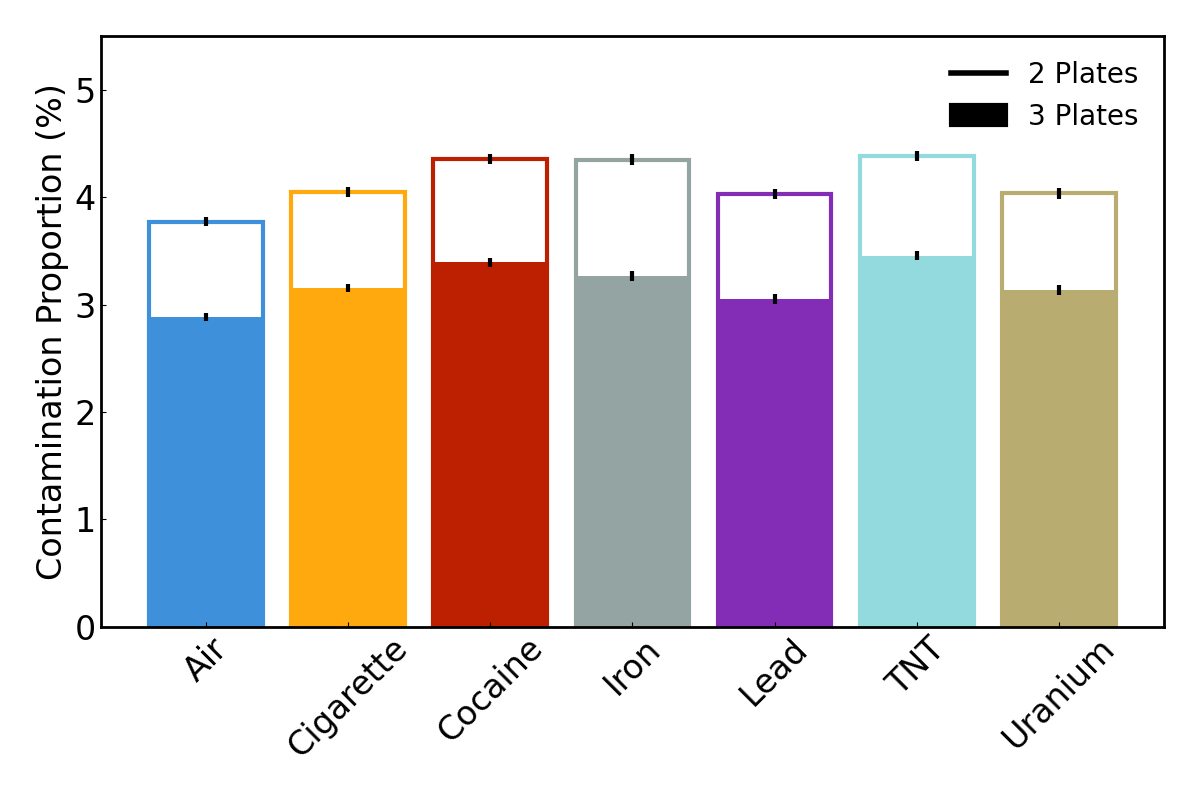}
    \caption{The contamination proportion for a variety of cargo types (each simulated as a 1\,m$\times$1\,m$\times$0.5\,m box centered in the container). The hodoscopes are simulated with either 2 (hollow bars) or 3 (solid bars) scintillator plates.}
    \label{fig:conf_twoplate}
\end{figure}

Fig.~\ref{fig:conf_twoplate} shows that whilst the contamination effect increases as expected, it is still at a very low level.\\

In conclusion, it is clear that with the current SilentBorder scanner setup the impact of secondary hits is very limited. The variation in the number of hits recorded when different cargoes are stored is too small to be used as a handle to investigate cargo compositions and the degradational impact on the muon tracking for MTS appears to also be very small. These studies are constrained to the impact of secondaries in a simple SilentBorder-like setup, there is significant other work in the realm of using secondary particles to add complementary information to muon tomography techniques. Many of these use specialized separate detector elements to better detect secondary particles [\onlinecite{instruments6040066, galgoczi2020imaging, osti_1097689, Bikit_2016}].

\section{TomOpt}
\label{sec:tomopt}
\subsection{TomOpt pipeline}
\begin{figure}[h!]
\includegraphics[width=\linewidth]{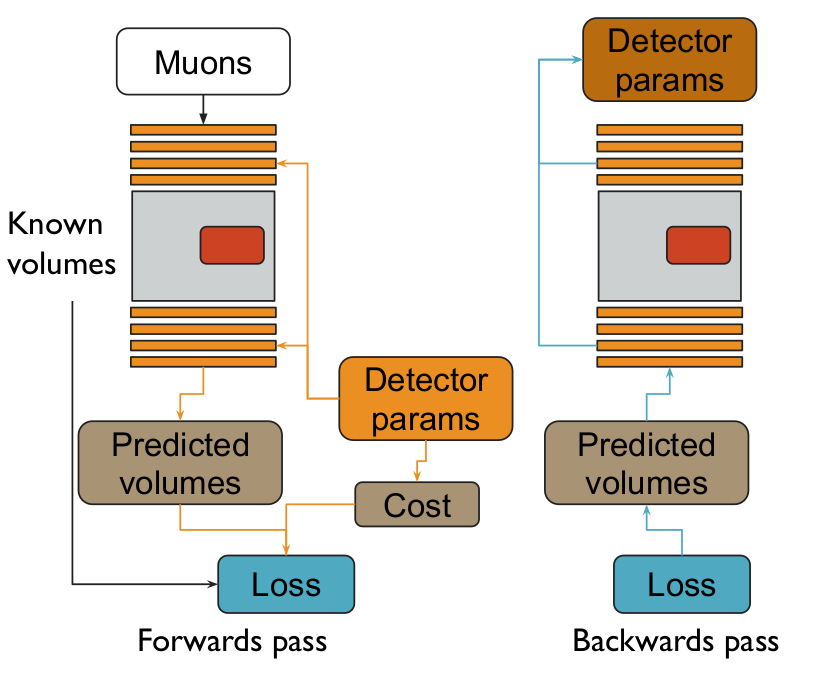}
\caption{TomOpt optimization pipeline, as described in [\onlinecite{TomOpt-Strong2023}]. Reproduced from Strong et al., TomOpt: Differential Optimisation for Task- and Constraint-Aware Design of Particle Detectors in the Context of Muon Tomography, Machine Learning: Science and Technology, 5(3), 035002 (2024); licensed under a Creative Commons Attribution (CCBY 4.0) license.}
\label{fig:tomopt}
\end{figure}

TomOpt [\onlinecite{TomOpt-Strong2023}] is a software tool developed for task-specific and constraint-aware optimization of the geometries and design specifications of detectors used for muon scattering tomography. Its modular Python [\onlinecite{python}] architecture features a fully differentiable, end-to-end pipeline that employs differentiable programming through utilizing the automatic differentiation capabilities of the leveraged PyTorch library [\onlinecite{pytorch}]. The optimization pipeline seamlessly integrates the various stages, including muon generation, muon detection and their propagation in scanned volumes leading to volume reconstruction, loss evaluation, and gradient backpropagation with respect to the detector design parameters. The detector parameters are then updated through gradient descent in the direction that minimizes the loss. A schematic representation of this pipeline is shown in Fig.~\ref{fig:tomopt}, illustrating the main stages in the process, which are further detailed below. The modularity of the simulation chain allows for tailored user-specific implementations of muon source and propagation models, detector parameterization, volume reconstruction methods, loss function definitions, etc.\\

\begin{figure}[h!]
\includegraphics[width=\linewidth]{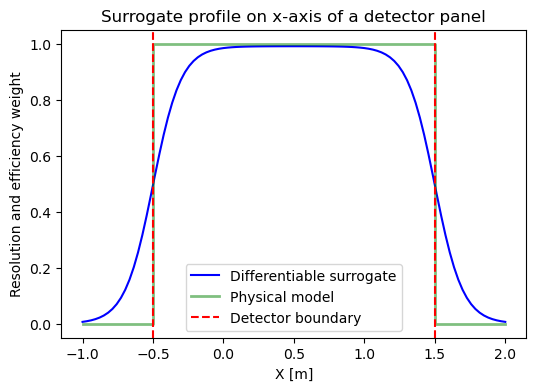}
\caption{Example of the surrogate model used to describe resolution and efficiency variations along the $x$-direction. For illustrative purposes, it is shown in one dimension (along the $x$-axis), but in practice, it is applied independently along both the $x$ and $y$ directions.}
\label{fig:res-wgt}
\end{figure}

Prior to optimization, detectors positioned above and below a volume of interest- referred to as the passive volume- are instantiated with sub-optimal parameters. These parameters can be the position in $xyz$, dimension, efficiency, etc. The passive volume is assigned a voxelized ground truth profile of radiation length ($X_0$). The optimization then runs over a specified number of epochs, where the detector parameters are updated at the end of each iteration. For each epoch, generated muons are sampled from defined flux models taken from literature [\onlinecite{guan2015}, \onlinecite{shukla2018}] and propagated through the detectors and passive volume setup. The scattering model reported by the Particle Data Group [\onlinecite{pdg}] is used for the simulation of the multiple Coulomb scatterings of muons in the passive volume and the calculation of their scattering angles and spatial displacements. The resolution and efficiency of the detectors are modeled by a differentiable surrogate function, represented as a double-sigmoid (Fig.~\ref{fig:res-wgt}). This function achieves maximum resolution and efficiency within the active area of the detector and smoothly decreases while transitioning outside the detector boundaries. Each muon hit in the detectors then is assigned an efficiency and a spatial resolution in $xy$ which is used to obtain the corresponding reconstructed hit with simulated detector response effect. This is done by sampling uncertainties in $x$ and $y$ from Gaussian distributions centered at the true position and of standard deviations equal to the corresponding spatial resolutions. Hits outside the detector region are still recorded but with degraded efficiency and resolution, which is a necessary feature to maintain the differentiability of the subsequent volume reconstruction with respect to the detector parameters. This allows for gradient descent to optimize the configuration which records the hits in an optimal way that yields best performance of the studied task. The hits above and below the passive volume are then fitted into incoming and outgoing tracks, respectively, using an analytical likelihood maximization method accounting for the uncertainties on the $x$ and $y$ positions. The Point-of-Closest-Approach algorithm is a common method used to reconstruct the scattering vertices and obtain a 3D density distribution of these vertices. This approach computes the closest point between the incoming and outgoing muon tracks in 3D space, which corresponds to the midpoint of the common perpendicular to the two tracks. The algorithm assumes that the entirety of the multiple Coulomb scatterings of a muon occurs at this reconstructed point. Subsequent inference methods utilize the spatial and/or angular variables of the reconstructed PoCA points to infer the physical properties of the passive volume under investigation. The discrepancy between the inferred, or predicted, passive volume quantities and their true values is then used to compute a task-specific loss function. Additionally, an optional term reflecting the budget constraints of the detector, calculated based on the unit surface area cost, may be included in the loss function. The loss is then back-propagated through the reconstruction pipeline, and its gradients are calculated with respect to the detector parameters. The detector optimization problem is formulated as a minimization task to be solved through gradient descent. This technique iteratively updates the detector parameters $a$ in the direction that minimizes the loss function: 

\begin{equation}
    a_{n+1} = a_n - \gamma \cdot \nabla_a \mathcal{L} (a_n),
\end{equation}

\noindent where $\gamma$ is a learning rate of the parameter and $\nabla_a 
 \mathcal{L} (a_n)$ is the gradient of the loss function with respect to $a$. This optimization cycle is repeated for multiple epochs until convergence is attained and the detector setup reaches its final optimized configuration.

\subsection{Volume reconstruction}

\subsubsection{Radiation length estimation \label{x0-section}}

The primary inference method in TomOpt estimates the radiation length ($X_0$) of the voxelized passive volume by leveraging reconstructed PoCA variables. Each PoCA vertex is modeled by a probability distribution function using three independent Gaussian distributions in the $x$, $y$, and $z$ dimensions, which expresses its probability to occur within a voxel. The Gaussian distributions $\mathcal{N}$ are characterized by their mean $\mu$ and standard deviation $\sigma$, where the mean corresponds to the nominal reconstructed position of the PoCA, and the standard deviation represents the uncertainty on that position within each respective dimension, computed by automatic differentiation: 
\begin{equation}
\text{PoCA pdf} = \prod_{i \in \{x, y, z\}} \mathcal{N}(x_i; \mu_i, \sigma_i)
\end{equation}

\begin{figure}
    \centering
    \includegraphics[width=0.5\textwidth]{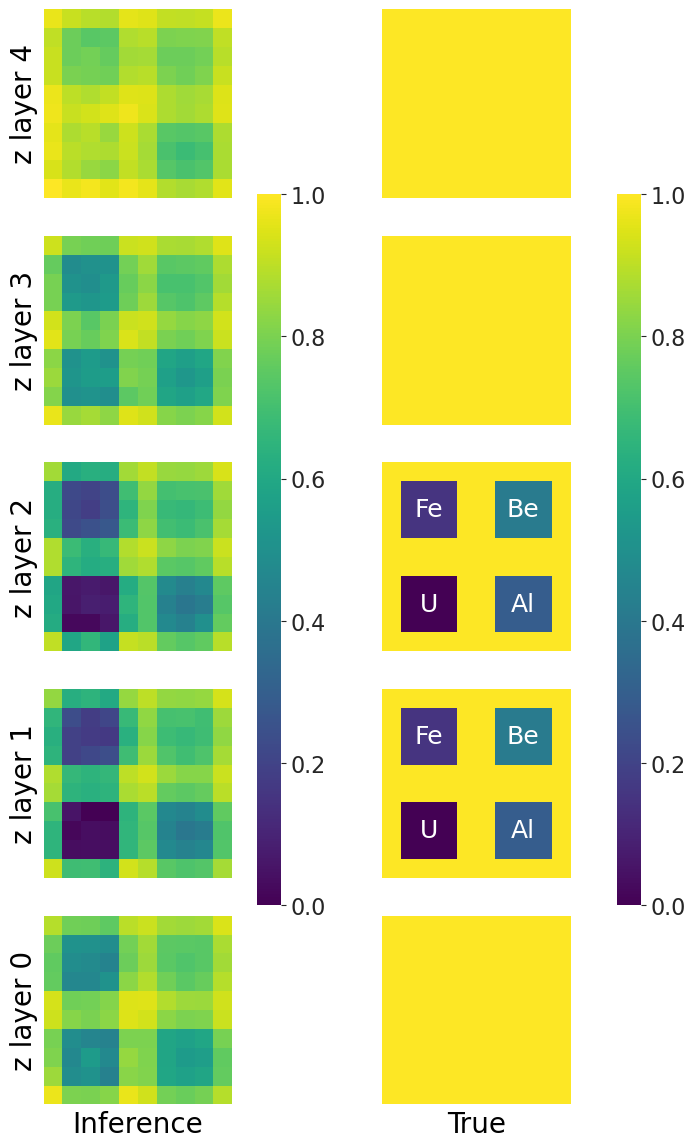}
    \caption{3D volume inference using the radiation length estimation method. The scanned volume consists of a $1~\unit{m} \times 1~\unit{m} \times 0.5~\unit{m}$ air-filled region, into which four material blocks \textemdash iron, uranium, aluminum, and beryllium\textemdash each measuring $30~\unit{cm} \times 30~\unit{cm} \times 20~\unit{cm}$, are inserted. These blocks span the second and third slices along the $z$-axis (denoted as $z$ layer 1 and $z$ layer 2). The right column presents cross-sectional heatmaps of the ground truth radiation lengths, while the left column shows the corresponding inferred values. Radiation lengths are displayed in a normalized natural logarithmic scale to enhance contrast between materials with similar densities.}
    \label{fig:x0-preds}
\end{figure}

The probabilistic modeling of PoCA vertices allows the influence of each reconstructed PoCA point to extend across multiple voxels within the passive volume. For each muon batch, each voxel is assigned a probability for every muon's PoCA point of its being located within that voxel. These probabilities are used as weights to compute a weighted root-mean-square (RMS) of the scattering angle distribution inside each voxel, taking into account all PoCA points in the muon batch. This RMS value is used to estimate the scattering behavior within that voxel. By inverting Eq.~\ref{eq:thetaRMS}, the RMS of the scattering angles is employed to infer the radiation length $X_0$ of the material in each voxel. Although the inferred radiation lengths exhibit a systematic bias toward lower values ([\onlinecite{TomOpt-Zaher2024}]), the resulting 3D image effectively distinguishes between high-$Z$ and low-$Z$ voxels, providing a clear representation of the material properties within the scanned volume (Fig.~\ref{fig:x0-preds}).

\subsubsection{A clustering density inference}

Another volume inference method that we adopt in TomOpt is based on a binned clustering algorithm [\onlinecite{bca}] that groups reconstructed scattering vertices, i.e. PoCA points in this case, into clusters indicating different materials in a voxelized volume. This clustering is done based on the spatial proximity and scattering angles of the scattering vertices. Since materials with different densities produce different scattering distributions, the properties of clustered PoCA points can reveal the density profile of the volume and determine high- and low-density regions.\\

\begin{figure}[h!]
\includegraphics[width=\linewidth]{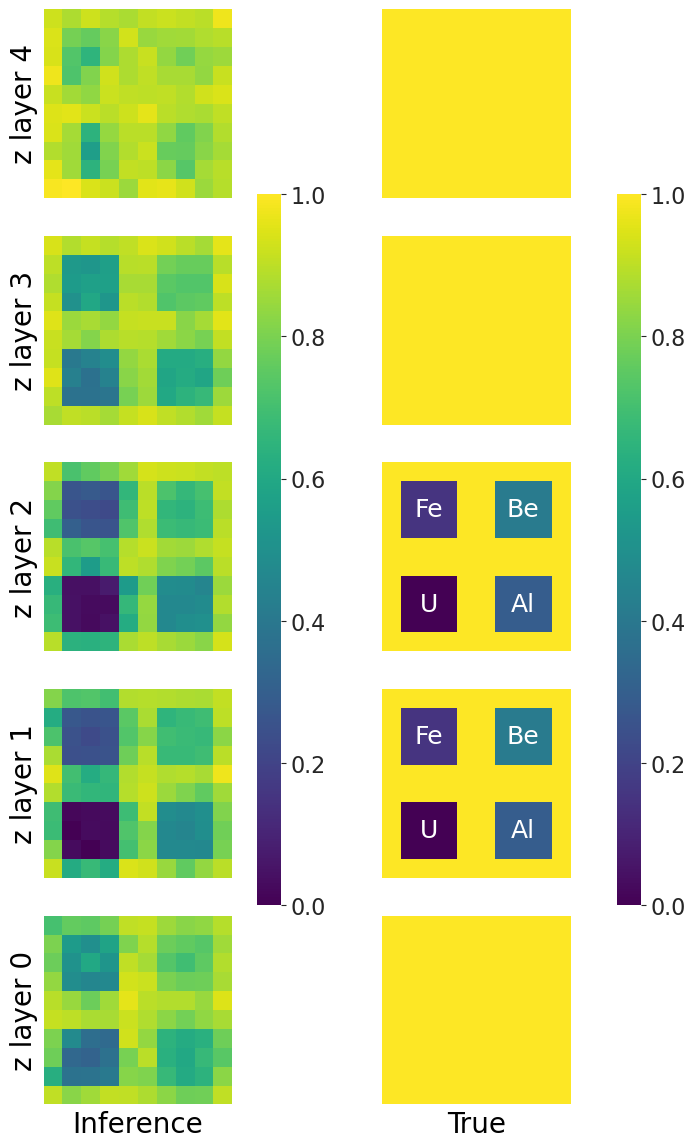}
\caption{3D volume inference using the binned clustering algorithm (BCA) method. The scanned volume consists of a $1~\unit{m} \times 1~\unit{m} \times 0.5~\unit{m}$ void (air), containing four inserted blocks of different materials: iron, uranium, aluminum, and beryllium. Each block measures $30~\unit{cm} \times 30~\unit{cm} \times 20~\unit{cm}$ and spans the second and third layers along the $z$-axis (labeled $z$ layer 1 and $z$ layer 2). The right column displays cross-sectional heatmaps of the ground truth radiation lengths in the $xy$-plane at each $z$ layer, visualized in a normalized natural logarithmic scale to enhance contrast between materials of similar density. The left column shows the corresponding inferred BCA voxel scores obtained using the BCA method, represented in a normalized linear scale. Note that the BCA algorithm inherently incorporates a logarithmic transformation, hence no additional log-scaling was applied to the inferred values.
}
\label{fig:bca-preds}
\end{figure}

Within each voxel, PoCA points are ranked in descending order of their scattering angles, and a fixed subset is selected to ensure uniform treatment across voxels. For each pair of selected PoCA points, their distance metric is computed and scaled by the inverse of the product of their scattering angles and normalized momenta. The normalized momentum is defined as 

\begin{equation}
    \tilde{p} = \frac{p}{p_{\text{norm}}},
\end{equation}

\noindent where \( p_{\text{norm}} \) is the nominal average cosmic muon momentum at sea level. Since most muon tomography setups lack direct momentum measurements for practical reasons, we assume a fixed momentum of 5 GeV for all muons, effectively setting \( \tilde{p} = 1 \). The median of the natural logarithm of the per voxel weighted metric distribution serves as an indicator of material density, with lower values corresponding to regions where scattering vertices are both larger and closer together. The natural logarithm is applied to improve the separation between closely clustered values. This median value defines the corresponding voxel score, resulting in a per volume distribution of voxel scores. The algorithm subsequently extracts the minimum score value, which serves as a discriminator for identifying the presence of a uranium block in the volume by applying a threshold. While this binary classification approach has been explored in previous work [\onlinecite{TomOpt-Zaher2025}], the present study shifts focus towards the 3D image reconstruction of the passive volume (Fig.~\ref{fig:bca-preds}), utilizing the full voxel score distribution. The previously mentioned extended PoCA pdf in Section~\ref{x0-section} is now used, eliminating the need to select a subset of the largest scattering vertices. This allows for a more efficient voxel population, reducing the number of muon events typically required to achieve uniform coverage.\\

\begin{figure}[h!]

    \centering
    \begin{subfigure}{\linewidth}
        \centering
        \includegraphics[width=\linewidth]{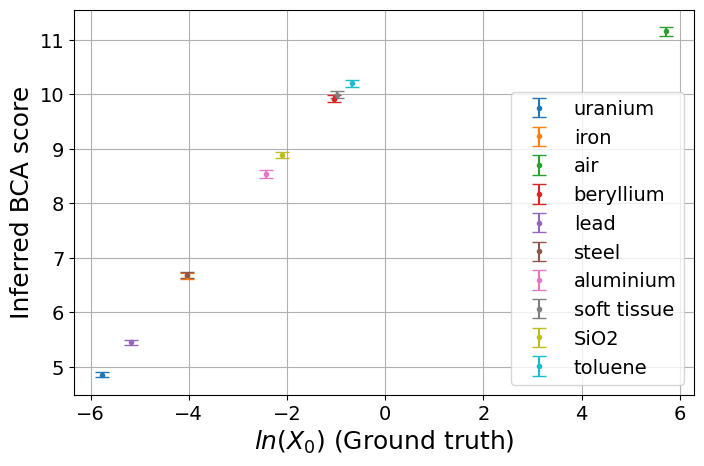}
        \caption{}
        \label{fig:bca-x0}
    \end{subfigure}%
    \\ 
    \begin{subfigure}{\linewidth}
        \centering
        \includegraphics[width=\linewidth]{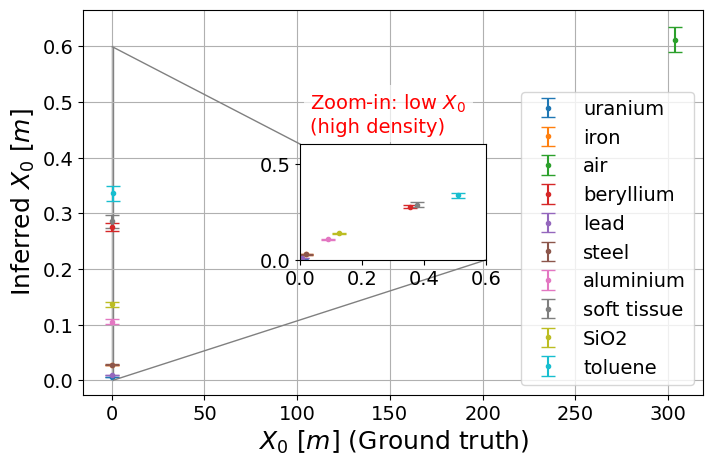}
        \caption{}
        \label{fig:x0-x0}
    \end{subfigure}
    \caption{Relationship between inferred quantities and ground truth $X_0$ for different materials. (a) shows the correlation between BCA voxel scores and the natural logarithm of the corresponding radiation length. A strong linear trend is observed for mid-to-high-Z materials, while the correlation weakens for low-Z materials such as air. The natural logarithm of $X_0$ is taken to ensure a valid comparison with the BCA score, as the latter inherently includes a logarithmic transformation in its metric. (b) compares inferred and true $X_0$ values, demonstrating a significant drop in inferred values observed for air, while higher-Z materials exhibit a near-linear relationship. A zoomed-in view reveals the enhanced linearity at higher Z.
}
    \label{fig:x0-relation}
\end{figure}

Fig.~\ref{fig:bca-x0} illustrates the relationship between the BCA voxel score and the ground truth value of the corresponding radiation length ($X_0$) for different materials, ranging from low to high Z materials. The statistics are derived from $300$ scans of a single voxel, distributed across the range of materials shown. The BCA score inference is performed using the extended PoCA version of the algorithm, with $1,000$ sampled muons. A strong linear correlation is evident when considering only the high-to-mid Z range, while this linearity breaks down for low Z materials, such as air. The natural logarithm of the $X_0$ values is used on the $x$-axis, as the BCA score calculation inherently involves the natural logarithm. In comparison, a similar plot (Fig.~\ref{fig:x0-x0}) showing the relationship between inferred and ground truth $x_0$ using the previous inference method reveals a significant drop in inferred values for air relative to the true value. This discrepancy underscores the limitations of the previous approach, particularly in accurately capturing low-Z materials. This relationship supports the use of the BCA inference as a reliable method for optimization, by minimizing the mean squared error (MSE) between the BCA predictions and the corresponding $X_0$ values (in natural logarithmic scale), both of which are normalized, to define the loss function.

\section{Bayesian optimization with TomOpt}
\label{sec:bo}
\subsection{The \texttt{Ax} library}

While TomOpt’s standard optimization approach relies on gradient descent, we investigate Bayesian Optimization (BO) [\onlinecite{Mockus1978}] as an alternative due to challenges posed by the volume inference methods. In particular, the presence of outliers in the inferred volume properties—especially in the $X_0$ inference approach—leads to a noisy mean squared error (MSE) loss function. Although gradient-based optimization is, in most cases, capable of improving detector coverage by optimizing the spatial positions of the hodoscopes in simple scenarios, such as configurations with a few hodoscopes, the sharp fluctuations in the MSE loss during optimization prevent the final configuration from being considered a fully optimized solution. This issue becomes more pronounced when extending the parameter space to include non-trivial design factors, such as the relative spacing of panels within the hodoscopes.\\

While ongoing efforts focus on mitigating inference biases, we present here a BO framework as a robust alternative for cases where the loss function is inherently noisy.\\

Bayesian Optimization (BO) is a powerful approach for optimizing black-box functions that are expensive to evaluate, lack a known analytical form or exhibit noise and uncertainty. Unlike gradient-based optimization methods, which require the objective function to be differentiable, BO constructs a probabilistic surrogate model of the objective function that approximates the function and guides the search for the optimal solution. At the core of BO is Bayes' theorem which states that - given prior knowledge about a function $f$ and newly observed data $\mathcal{D}$, the posterior distribution of $f$ is proportional to the likelihood of $\mathcal{D}$ given $f$ multiplied by the prior distribution of $f$ [\onlinecite{Brochu2010}]:

\begin{equation}
    p(f \mid \mathcal{D}) \propto p(\mathcal{D} \mid f) p(f).
\end{equation}

A common choice for the surrogate model in BO is a Gaussian Process (GP), a flexible, non-parametric model that provides a probabilistic estimate of the objective function. A GP assumes that the function values at different points are correlated and defines a prior distribution over possible functions. Given a set of observed data points, the GP updates its belief about the function by computing a posterior distribution, which provides both a mean prediction (expected function value) and a variance estimate (uncertainty in the prediction). This uncertainty measure is particularly advantageous in BO, as it enables informed decisions about where to evaluate the function next. The selection of evaluation points is determined by an acquisition function, which defines a utility function from the model posterior that determines the next point to evaluate [\onlinecite{Brochu2010}]. The selection is performed in a way that strategically balances exploration (sampling uncertain regions) and exploitation (sampling regions with promising performance). 
One widely used acquisition function is Expected Improvement (EI), which quantifies how much improvement we expect by sampling at a particular point. In the case of minimizing the objective function, the improvement at a new point $x$ is defined as the positive difference between the current best observed minimum $f^*$ and the predicted function value $f(x)$, if the new point offers an improvement:

\begin{equation}
    I(x) = \max(f^* - f(x), 0)
\end{equation}

\noindent where  $f^* = \min_{i=1,\dots,n} f(x_i)$ is the lowest observed function value so far and $f(x)$ is the function value at the candidate point $x$. The EI is then computed by taking the expectation of \( I(x) \) over the posterior distribution of \( f(x) \), modeled by the GP [\onlinecite{Jones1998}]:
\begin{equation}
    \mathbb{E}[I(x)] = \mathbb{E} \left[ \max(f^* - f(x), 0) \right].
\end{equation}

A higher EI indicates that sampling at that point is expected to yield a larger improvement over the current best, making it a good candidate for the next evaluation. After evaluating a new point, the surrogate model is updated iteratively, refining the search for the global optimum while minimizing the number of function evaluations.\\

\texttt{Ax} [\onlinecite{Ax}] is an open-source platform for adaptive experimentation and Bayesian Optimization, developed by Meta. It provides a high-level interface for optimizing parameters without requiring explicit definition of the acquisition function. \texttt{Ax} internally leverages BoTorch [\onlinecite{botorch}], a library for Bayesian Optimization, and automatically selects an acquisition function based on the optimization scenario.\\

In this work, BO is implemented using the \texttt{Ax} Loop API, which  provides an automated framework for sequential BO. The optimization pipeline consists of defining the search space, constructing and updating the surrogate model, and iteratively selecting and evaluating parameter configurations. The search space defines the range of possible values for the parameters being optimized, in our case the $xyz$ positions of hodoscopes. This parameter space is passed to the experiment object, which acts as the central controller for the optimization process. The experiment object in \texttt{Ax} manages the entire optimization process, including the collection of observations, selection of new trials, and updating of the surrogate model. It stores all configurations tested so far along with their corresponding objective function values. Evaluation of each candidate set of parameters is done through a user-defined evaluation function that quantified the performance of the parameters. We take the MSE of the volume inference as our evaluation function, which quantifies the accuracy of the voxelized reconstructed material distribution and directly relates to the ability to discriminate high-density or suspicious objects in border security applications. BO with \texttt{Ax}'s Loop API is conducted using a hybrid sampling strategy that consists of two sequential phases: 
\begin{itemize}
    \item Sobol generation strategy: \\
    The first 12 trials are evaluated using a low-discrepancy quasi-random Sobol sequence that ensures uniform coverage of the search space. These initial evaluations serve to bootstrap the Gaussian Process model by providing an initial dataset that prevents premature convergence to a local optimum. Sobol sampling is preferred over purely random initialization as it minimizes clustering and ensures better space-filling properties.

    \item Gaussian Process Regression (GPEI):\\
    After the initial Sobol trials, \texttt{Ax} switches to Bayesian Optimization generation strategy using a Gaussian Process surrogate model prior of the objective function coupled with an Expected Improvement acquisition function (GPEI).
\end{itemize}

This hybrid strategy ensures efficient exploration of the parameter space while minimizing the number of function evaluations required for convergence. At the end of the optimization process, the best parameter configuration is extracted based on the lowest observed objective function value.\\

For our purpose, we integrate \texttt{Ax}’s BO framework with TomOpt, coupling the BO optimization pipeline to the TomOpt simulation pipeline while bypassing the gradient-based optimization module. Instead of using gradient descent to optimize detector parameters, we employ \texttt{Ax}’s BO strategy, where each evaluation trial consists of updating the current detector parameter set, running a full muon tomography simulation with the updated detector configuration, and performing volume inference. These inferred values are then used to evaluate the surrogate objective function through a MSE evaluation function, providing feedback for the next trial.\\

Since the MSE loss function is sensitive to small errors and exhibits noise, BO is particularly useful for optimizing our detector parameters. The surrogate model in BO allows us to smooth out fluctuations in the loss landscape, leading to more reliable parameter tuning. Additionally, using BO enables a controlled comparison of our two inference methods: radiation length inference and the binned clustering algorithm. By setting aside gradient-based updates, we can assess how each inference method performs under the same optimization framework, providing insights into their stability and effectiveness in reconstructing material properties from muon scattering data.\\

\begin{figure}[ht]
    \centering
    \includegraphics[width=\linewidth]{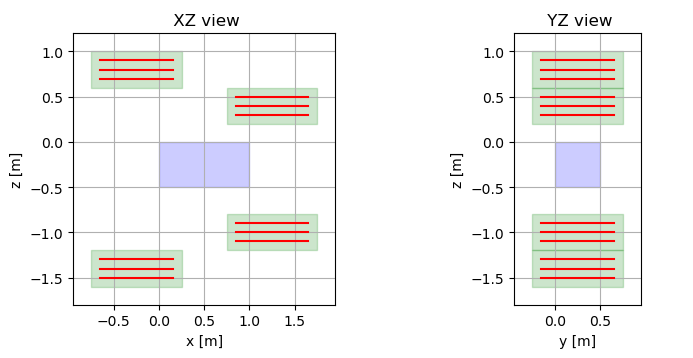}
    \caption{The initial configuration of the hodoscope setup in XZ (left) and YZ (right) views.}
    \label{fig:init-hods}
\end{figure}

We use a passive volume with dimensions $1~\unit{m} \times 1~\unit{m} \times 0.5~\unit{m}$, simulated as a lorry with an iron base and filled with a random arrangement of $10~\unit{cm} \times 10~\unit{cm} \times 10~\unit{cm}$ iron, beryllium and air blocks to increase scenario variability and improve the generalizability of the results. Additionally, a $2000~\unit{cm^3}$ uranium block is randomly placed inside the lorry with a probability of 0.5 of being present. The initial detector setup consists of a hodoscope pair, each measuring $1.5~\unit{m} \times 1.5~\unit{m} \times 0.4~\unit{m}$, placed above and below the passive volume. The maximum of the surrogate for spatial resolution is 0.1~\unit{mm}, corresponding to the uncertainty on a single muon hit position, as in TomOpt each muon is generated and processed individually. With current computational resources, TomOpt comfortably handles objects of this scale. For significantly larger volumes, longer exposure times are required, translating into a larger number of simulated muons and correspondingly bigger PyTorch tensors storing both muon and voxel information. This can become a bottleneck if computations are run solely on CPU. While GPU acceleration can alleviate part of this burden, scaling to much larger objects (e.g. tens of meters) remains a practical challenge rather than a conceptual one.
Initially, hodoscopes are offset by $0.5~\unit{m}$ in both $x$ and $y$ directions, leaving part of the passive volume uncovered (Fig.~\ref{fig:init-hods}).\\

The reason for placing the hodoscopes at different $z$ positions is to avoid overlap along the $z$ axis. This is due to the current implementation of hodoscopes in TomOpt, where each hodoscope is defined as a single unit consisting of individual detector plates. The fitting of hits into tracks accounts for hits in all detector plates, where hits outside an active plate surface are down-weighted  through lower resolution and efficiency. Hence, placing two plates at the same $z$ level would cause ambiguities in track reconstruction. While this constraint does not prevent the optimisation study focusig on inference methods, it does limit the ability to model perfectly co-planar panels or more complex detector layouts where modules intentionally share the same z-position.This panel overlap limitation is a technical constraint of the current version and is being addressed in an updated implementation that will be presented in a forthcoming paper.\\

 Starting from this suboptimal configuration, the optimization is expected to converge toward an optimal final setup where the hodoscopes perfectly cover the passive volume. BO is then executed for $100$ evaluation trials. Before each evaluation, the detector parameters are set to the values sampled for evaluation.

\subsection{Optimization with BCA inference method}

\begin{figure}[h!]
    \centering
    \includegraphics[width=1.1\linewidth]{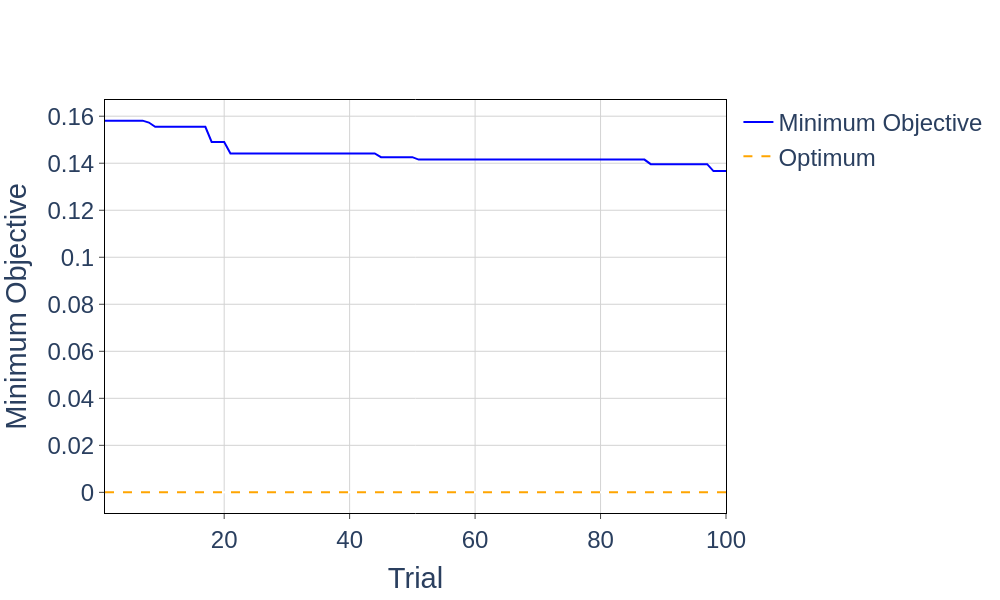}
    \caption{Convergence of the best minimum objective (ideal optimum = 0) across 100 BO trials using BCA inference.}
    \label{fig:best-obj-bca}
\end{figure}

In this section, BO is conducted using the BCA inference method for volume reconstruction. The surrogate objective function is evaluated at each trial by computing the MSE between inferred BCA voxel scores and the logarithmic ground truth $X_0$, both normalized to unity. This formulation captures the underlying correlation between the two quantities, as described in Fig.~\ref{fig:bca-x0}.\\
Fig.~\ref{fig:best-obj-bca} shows the convergence of the best minimum objective over 100 trials on CPU,  showing a 13.3\% reduction throughout the optimization. For simplicity, only a single run is shown, although different random seeds may yield different results. The full optimization takes $\sim$26 minutes, while using gradient descent would result in $\sim$38\% increase in runtime. This difference arises because the BCA reconstruction is computationally heavier than the formula-based $X_0$ inference method, involving PoCA pair-wise voxel clustering. The gradient descent optimization requires evaluating many gradients per step, resulting in longer runtime. With BO, the GP model reduces the number of expensive evaluations, so it is faster overall despite the overhead of building the GP model.
 
\begin{figure}[h!]

    \centering
    \begin{subfigure}{\linewidth}
        \centering
        \includegraphics[width=\textwidth]{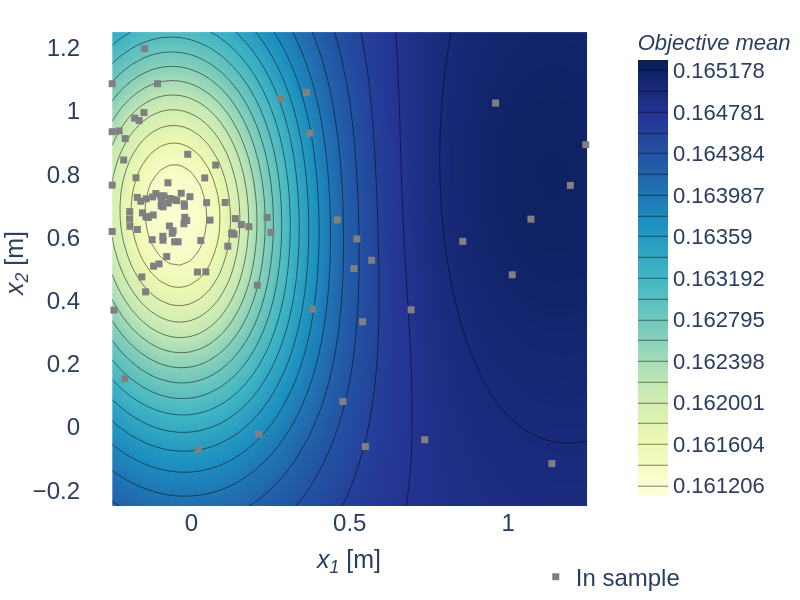}
\caption{}
        \label{fig:mean-x0}
    \end{subfigure}%
    \\ 
    \begin{subfigure}{\linewidth}
        \centering
        \includegraphics[width=\linewidth]{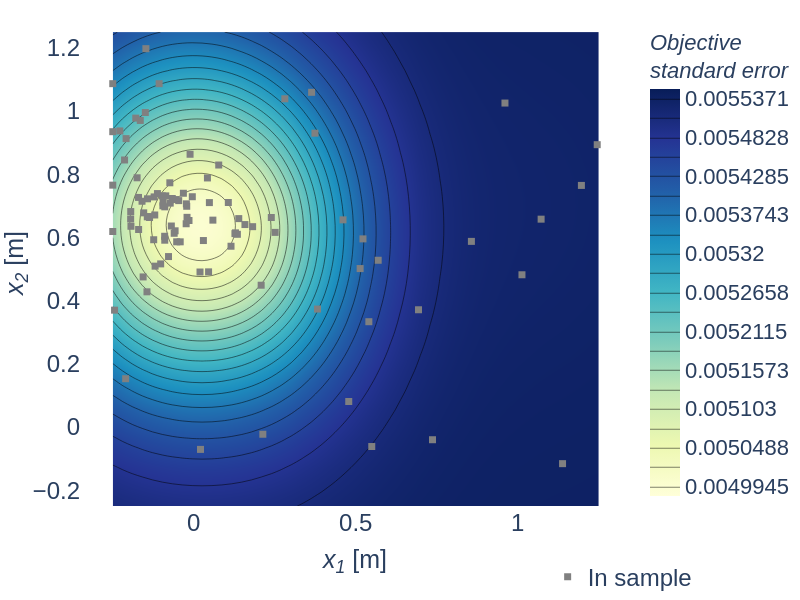}
        \caption{}
        \label{fig:error-x0}
    \end{subfigure}
    \caption{Response surface of the objective mean (a) and standard error (b) for the $x$ positions of the upper two hodoscopes, labeled $x_1$ and $x_2$. The objective function being evaluated is a MSE that captures the correlation between inferred BCA voxel score and true voxel $X_0$ of the scanned volume. }
    \label{fig:contour-bca}
\end{figure}

Fig.~\ref{fig:contour-bca} presents the response surface for the $x$ positions of the two upper hodoscopes, illustrating both the mean and standard error of the objective function's predictive model. The response surface provides insight into how the objective function varies across the parameter space and highlights regions of lower expected error. 

\begin{figure}[h!]

    \centering
    \begin{subfigure}{\linewidth}
        \centering
        \includegraphics[width=\linewidth]{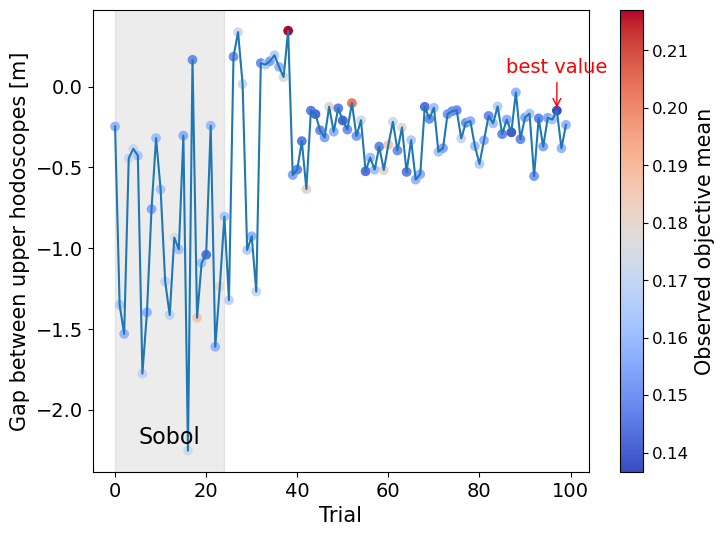}
\caption{}
        \label{}
    \end{subfigure}%
    \\ 
    \begin{subfigure}{\linewidth}
        \centering
        \includegraphics[width=\linewidth]{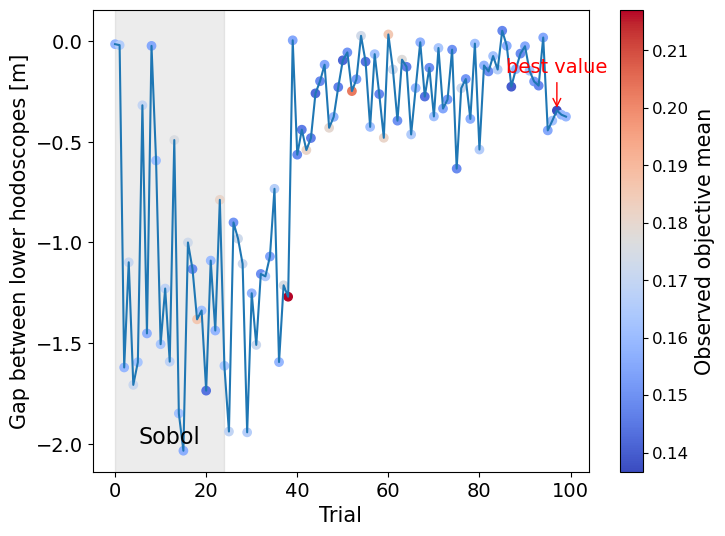}
        \caption{}
        \label{}
    \end{subfigure}
    \caption{The evolution of the gap size between upper hodoscopes (a) and between lower hodoscopes (b) across 100 trials during the BO process utilizing the BCA inference method. The color scale of the markers corresponds to the observed (evaluated) mean of the objective. For the first 24 trials, sampled using the Sobol strategy, a broader range of gap sizes is explored. Negative values indicate an overlap or swapping of the hodoscope pair along the x-direction. A trend is observed for the upper gap, where the gap size tends toward zero, suggesting minimal overlap. The lower gap shows a preference for overlap roughly equivalent to half the size of a hodoscope. The best gap values, corresponding to the optimal parameter configurations, are annotated in the plot. Notably, these values align with the minimum observed mean of the objective function.}
    \label{fig:gaps-bca}
\end{figure}

\begin{figure}[h!]
    \centering
    \includegraphics[width=1.1\linewidth]{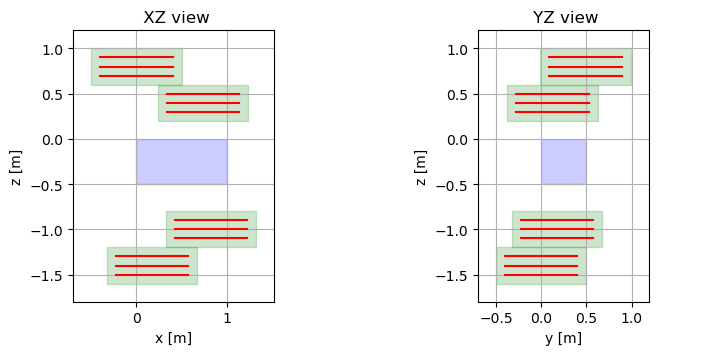}
    \caption{The BO optimized configuration of the hodoscope setup in XZ (left)and YZ (right) views. The placement of the hodoscopes aligns with the observations in Fig.~\ref{fig:gaps-bca}, where the upper hodoscopes provide fuller coverage, while the lower hodoscopes exhibit overlap while maintaining sufficient coverage of the passive volume.}
    \label{fig:final-bca}
\end{figure}

Fig.~\ref{fig:gaps-bca} displays the evolution of the gap size between the upper hodoscopes, as determined by the evaluated $x$ coordinates throughout the optimization trials. The marker color scale in this figure represents the observed (evaluated) mean of the objective function. The results demonstrate a clear trend in the optimization process, with BO guiding the hodoscope placement toward regions associated with lower objective values. The response surface and gap evolution plots collectively validate the surrogate model's effectiveness in navigating the noisy objective landscape. Fig.~\ref{fig:final-bca} presents the final optimized configuration of the detector setup, corresponding to the optimal parameter values identified by the BO process.

\subsection{Optimization with $X_0$ inference method}

\begin{figure}[h!]
    \centering
    \includegraphics[width=\linewidth]{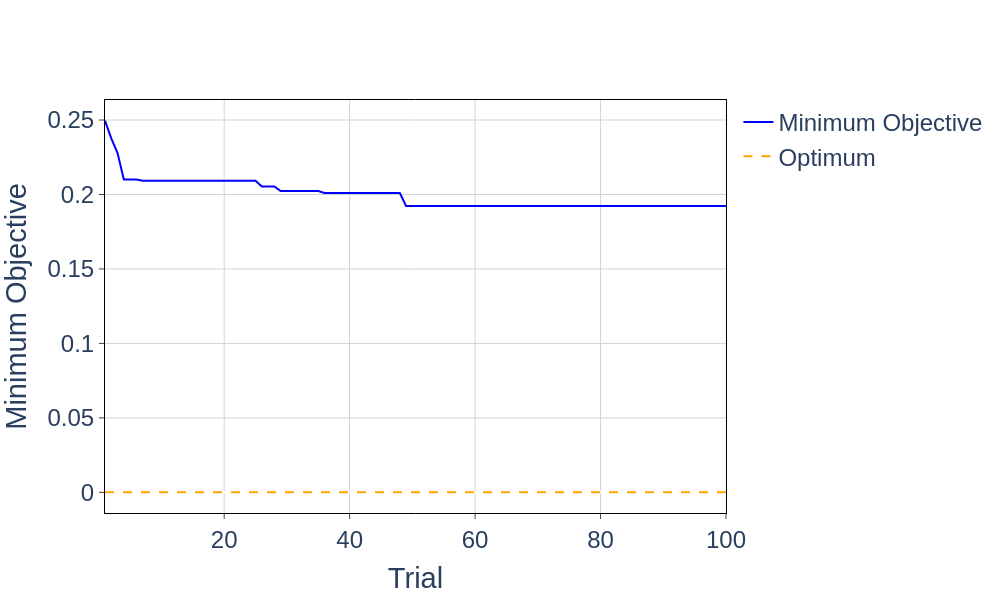}
    \caption{Convergence of the best minimum objective across 100 BO trials using $X_0$ inference.}
    \label{fig:best-obj-x0}
\end{figure}

\begin{figure}

    \centering
    \begin{subfigure}{\linewidth}
        \centering
        \includegraphics[width=\linewidth]{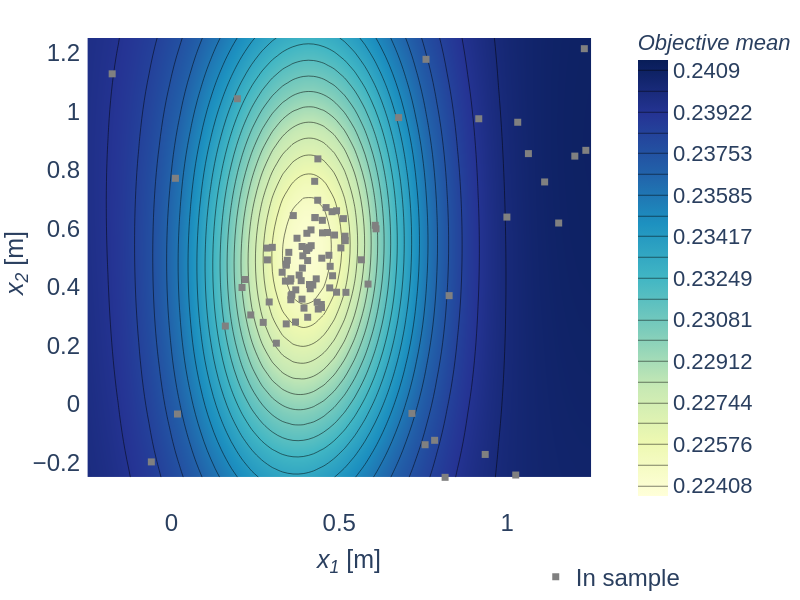}
\caption{}
        \label{fig:mean-x0}
    \end{subfigure}%
    \\ 
    \begin{subfigure}{\linewidth}
        \centering
        \includegraphics[width=\linewidth]{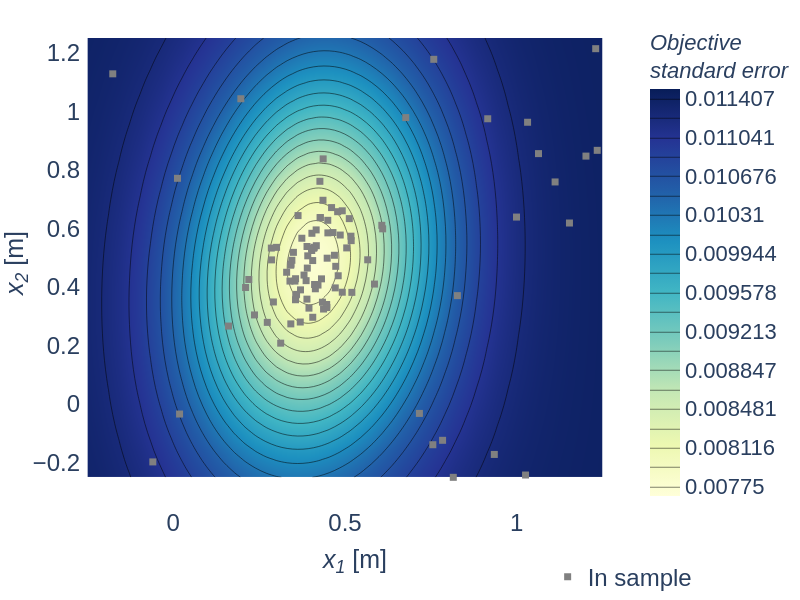}
        \caption{}
        \label{fig:error-x0}
    \end{subfigure}
    \caption{Response surface of the objective mean (a) and standard error (b) for the $x$ positions of the upper two hodoscopes, labeled $x_1$ and $x_2$. The objective function being evaluated is the MSE between the normalized inferred and true voxel $X_0$ values of the scanned volume. }
    \label{fig:contour-x0}
\end{figure}

\begin{figure}[h!]

    \centering
    \begin{subfigure}{\linewidth}
        \centering
        \includegraphics[width=\linewidth]{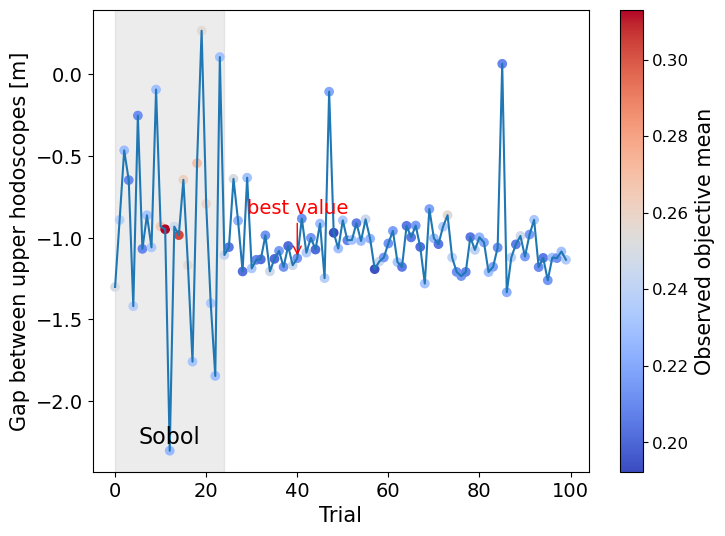}
\caption{}
        \label{}
    \end{subfigure}%
    \\ 
    \begin{subfigure}{\linewidth}
        \centering
        \includegraphics[width=\linewidth]{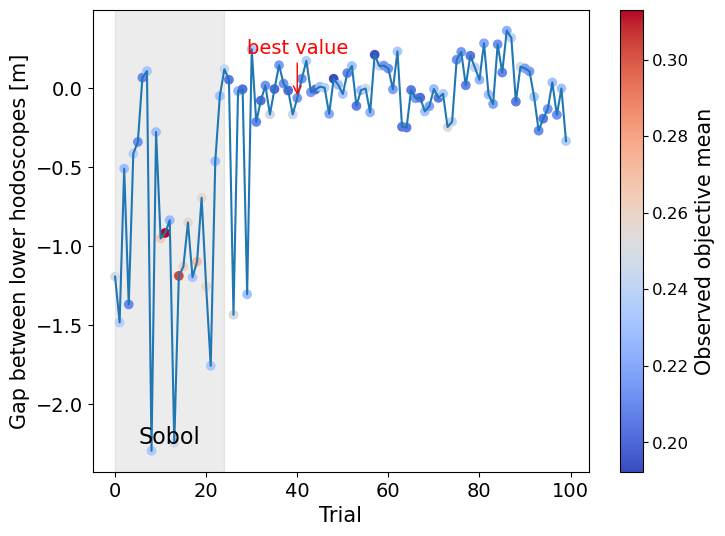}
        \caption{}
        \label{}
    \end{subfigure}
    \caption{The evolution of the gap size between upper hodoscopes (a) and between lower hodoscopes (b) across 100 trials during the BO process utilizing the $X_0$ inference method. The color scale of the markers corresponds to the observed (evaluated) mean of the objective. For the first 24 trials, sampled using the Sobol strategy, a broader scope of gap sizes is explored. Negative values indicate an overlap or swapping of the hodoscope pair along the x-direction. Unlike the results obtained using BCA inference, the optimal parameters are not located near the final trials but instead appear in the mid-trial range. Additionally, the various trial evaluations yield comparable values of the observed objective mean, as indicated by the predominance of blue-colored markers throughout the GPEI stage following Sobol sampling.}
    \label{fig:gaps-x0}
\end{figure}

In this approach, BO was performed using the $X_0$ inference method, where the evaluation function was defined as the mean squared error (MSE) between normalized inferred and ground truth $X_0$ values. Normalization was necessary due to the significant underbias observed for low-Z materials, particularly air (Fig.~\ref{fig:x0-x0}). Since air is present in the passive volume, the raw MSE values would otherwise be orders of magnitude larger than what is expected from a typical MSE evaluation, potentially distorting the optimization process. Fig.~\ref{fig:best-obj-x0} shows the convergence of the best minimum objective, with a total reduction of 22.8\%. The complete optimization runs $\sim$11 minutes on CPU with BO, comparable to $\sim$9 minutes with gradient descent. Here, the formula-based $X_0$ inference method is relatively computationally inexpensive, so the additional runtime in BO stems primarily from the overhead of updating the GP surrogate model.
 Fig.~\ref{fig:contour-x0} illustrates the response surface for the mean and standard error of the surrogate objective. Unlike the case with BCA inference, the gap sizes between the upper and lower hodoscope pairs (Fig.~\ref{fig:gaps-x0}) do not exhibit a clear optimization trend, with the best-evaluated parameters appearing mid-range in the trials and the observed mean of the objective function remaining relatively stable throughout the GPEI stage following Sobol sampling. This behavior can be attributed to the noisiness of the $X_0$ MSE, which arises from the inherent bias in the $X_0$ inference method.\\

\begin{figure}[h!]
    \centering
    \includegraphics[width=\linewidth]{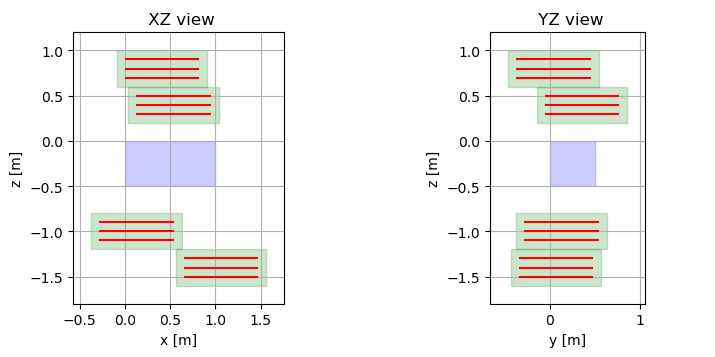}
    \caption{The BO optimized configuration of the hodoscope setup in XZ (left) and YZ (right) views. }
    \label{fig:final-x0}
\end{figure}

Although the $X_0$ inference approach still led to an optimal final detector configuration (Fig.~\ref{fig:final-x0}), achieving full coverage placement, its noisier optimization landscape suggests that it may not be as reliable for extending BO to more complex optimization problems. If the goal is to optimize additional detector parameters beyond basic placement, using a more stable inference method, such as BCA, may be necessary for efficient and interpretable optimization.

\begin{figure}[h!]
    \centering
    \includegraphics[width=\linewidth]{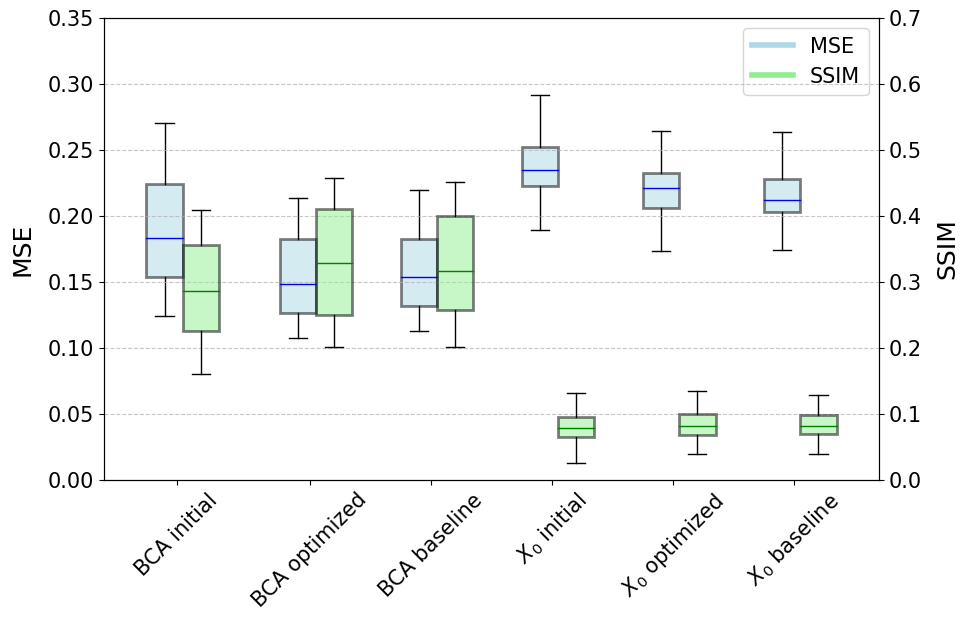}
    \caption{Statistical distribution of volume inference performance using MSE (left axis) and SSIM (right axis), across the initial, optimized and baseline detector configurations with the BCA and $X_0$ inference methods. Each boxplot summarizes the distribution of inference quality over 100 simulated passive volumes: 50 volumes containing blocks of iron, beryllium, and air, and 50 volumes with an additional randomly placed uranium block. In each boxplot, the central line indicates the median; the box edges represent the first and third quartiles (Q1 and Q3); and the whiskers extend to the most extreme data points within 1.5 times the inter-quartile range (IQR). For SSIM, higher values (closer to 1) indicate better structural similarity, whereas for MSE, lower values indicate better inference accuracy. Using the radiation length estimation method, a slight improvement is observed in MSE with the optimized configuration, while SSIM remains similar across all configurations. In contrast, the BCA method shows consistent improvement in both MSE and SSIM when using the optimized detector design, comparable to the baseline detector. }
    \label{fig:metrics}
\end{figure}

\begin{figure}[h!]
    \centering
    \includegraphics[width=\linewidth]{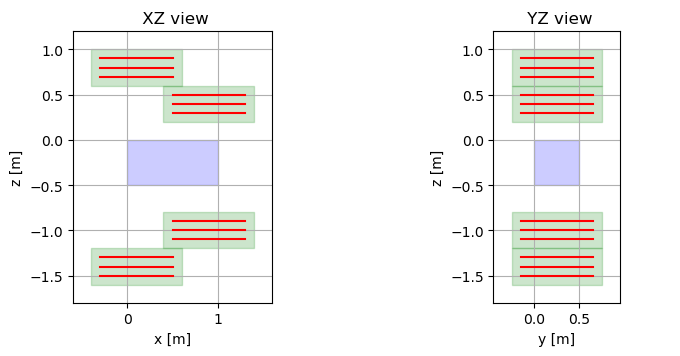}
    \caption{The configuration of the baseline detector setup in XZ (left) and YZ (right) views. }
    \label{fig:baseline}
\end{figure}

To statistically compare the inference performance of the optimized detector configurations, we evaluate the Mean Squared Error (MSE) and the Structural Similarity Index Measure (SSIM) for 100 simulated passive volumes, with and without a high-density uranium block (Fig.~\ref{fig:metrics}). The performance is compared with a baseline corresponding to a detector design based on human intuition for maximizing acceptance (Fig.~\ref{fig:baseline}). While the MSE directly quantifies voxel-wise differences, the SSIM is a perceptual metric that evaluates the structural similarity between predicted and true volumes. It is computed over local regions using a sliding window of size $4 \times 4$, and the final SSIM score is obtained by averaging over all such windows. The SSIM is defined as:

\begin{align}
\text{SSIM}&(infer,~true) =\nonumber \\ &\frac{(2\mu_{infer}\mu_{true} + C_1)(2\sigma_{infer,~true} + C_2)}{(\mu_{infer}^2 + \mu_{true}^2 + C_1)(\sigma_{infer}^2 + \sigma_{true}^2 + C_2)}
\end{align}

\noindent where $\mu_{infer}$ and $\mu_{true}$ are the local means, $\sigma_{infer}^2$ and $\sigma_{true}^2$ are the local variances, and $\sigma_{infer,~true}$ is the local covariance of the inferred and ground truth volumes, where the inferred quantities may represent radiation lengths or BCA scores depending on the inference method, and the ground truth is the true $X_0$ layout of the volume. The constants $C_1$ and $C_2$ are used to stabilize the division when the denominators are small. Higher SSIM values (closer to 1) indicate stronger structural agreement, while lower MSE values indicate higher accuracy. Overall, the BCA method exhibits superior inference performance compared to the radiation length estimation ($X_0$) method, both in terms of SSIM and MSE. This suggests that the BCA method is more robust to noise in complex or high-contrast configurations.\\

Some of the remaining limitations of TomOpt arise from intrinsic challenges in muography: the inverse problem is inherently ill-posed and relies on approximations, with the PoCA method reducing multiple scattering to a single point and introducing systematic biases. Importantly, nearly all current reconstruction methods build on PoCA, meaning that the sensitivity of detector design to volume reconstruction is closely tied to the accuracy of this approximation, highlighting the importance of developing approaches that go beyond PoCA in future work.

\section{Conclusions}
\label{sec:conclusion}

This work outlines a clear pathway toward the optimization of a future SilentBorder scanner for muon scattering tomography, with current results demonstrating meaningful advances in the studied approaches. Future studies will aim to extend these developments toward more realistic operational scenarios and validation with experimental data. Studies using the GEANT4 framework have shown that minimizing the in-plane gaps between active hodoscope areas, by keeping any casing and electronics sizes as small as possible, will ensure optimal efficiency of the system. However, increasing the vertical spacing between hodoscope plates does show improvement in the angular resolution with little reduction in efficiency and is worth pursuing. A complementary GEANT4 study on the role of secondary particle hits in event reconstruction was also conducted. Our findings suggest that secondary particles do not provide any extra material discrimination power in a `SilentBorder' type scanner but also cause very limited degradation of the reconstruction quality and therefore do not need separate measures to exclude them.\\

On the machine learning side, a Bayesian optimization (BO) module was developed within the TomOpt framework to enable robust design optimization for reconstruction methods affected by noise. This module complements the existing gradient-based approach and is particularly well-suited for scenarios where the loss landscape is poorly defined or convergence is difficult to achieve. The BO strategy was applied to two distinct reconstruction algorithms: a root-mean-square (RMS)-based $X_0$ inferrer and a binned clustering-based density inferrer. Results indicate that the RMS-based method is more sensitive to noise, whereas the binned clustering approach exhibits greater robustness. Although BO-optimized configurations demonstrated performance improvements, they did not significantly outperform a baseline detector configuration derived from human intuition. 
This motivates further investigation of deep learning–based approaches for volume inference to enhance sensitivity to subtle updates of detector parameters. This could be pursued either by coupling existing reconstruction outputs with post-processing neural networks, or by developing fully end-to-end learning frameworks that directly infer volumes from minimally processed detector-level scattering information.

\clearpage

\begin{acknowledgments}
The authors acknowledge funding from the EU Horizon 2020 Research and Innovation Programme under grant agreement no. 101021812 (“SilentBorder”) and by the Fonds de la Recherche Scientifique - FNRS under Grants No. T.0099.19 and J.0070.21.
Pietro Vischia's work was supported by the ``Ram\'on y Cajal” program under Project No. RYC2021-033305-I funded by the MCIN MCIN/AEI/10.13039/501100011033 and by the European Union NextGenerationEU/PRTR. Finally, the authors gratefully acknowledge the computer resources at Artemisa, funded by the European Union ERDF and Comunitat Valenciana as well as the technical support provided by the Instituto de Fisica Corpuscular, IFIC (CSICUV).
\end{acknowledgments}

\section*{Data Availability}
The data that support the findings of this study are available from the corresponding author upon reasonable request.

\bibliography{bibliography}

\end{document}